\documentclass[]{weauth}

\usepackage{moreverb}

\usepackage{graphicx}
\usepackage{epstopdf}
\usepackage{natbib}
\usepackage{multicol}
\usepackage{grffile}  
\usepackage{color}

\usepackage[colorlinks,bookmarksopen,bookmarksnumbered,
citecolor=red,urlcolor=red]{hyperref}

\newcommand\BibTeX{{\rmfamily B\kern-.05em \textsc{i\kern-.025em b}\kern-.08em
T\kern-.1667em\lower.7ex\hbox{E}\kern-.125emX}}

\def\volumeyear{2015}

\DeclareMathOperator\erf{erf}

\newcommand\Real{\mbox{Re}} 
\newcommand\Imag{\mbox{Im}} 
\newcommand{\ihat}{\boldsymbol{\hat{{i}}}}
\newcommand{\jhat}{\boldsymbol{\hat{{j}}}}

\newcommand{\colorTony}{black}

\begin{document}

\runningheads{Mart\'inez-Tossas, Churchfield
	and Meneveau}{}

\articletype{RESEARCH ARTICLE}

\title
{Optimal smoothing length scale for actuator line models of wind turbine blades}

\author{L.A. Mart\'inez-Tossas\affil{1},
M.J. Churchfield\affil{2},
and C. Meneveau\affil{1}}

\address{\affilnum{1}Department of Mechanical Engineering, Johns Hopkins University, 
Baltimore, MD 21218, USA \\
{\color{\colorTony}
\affilnum{2}National Renewable Energy Laboratory, 
Golden, CO 
}}

\corraddr{Journals Production Department, John Wiley \& Sons, Ltd,
The Atrium, Southern Gate, Chichester, West Sussex, PO19~8SQ, UK.}

\begin{abstract}
The actuator line model (ALM) is a commonly used method to represent lifting surfaces such as wind turbine blades within Large-Eddy Simulations (LES).
In the ALM the lift and drag forces are replaced by an imposed body force which is typically smoothed over several grid points using a Gaussian kernel  
with some prescribed smoothing width $\epsilon$.  To date, the choice of  $\epsilon$ has most often been based on numerical considerations related to the grid spacing used in LES.  However, especially for finely resolved LES with grid spacings on the order of or smaller than the chord-length of the blade, the best choice of  $\epsilon$ is not known.
In this work, a theoretical approach is followed to determine the most suitable value of $\epsilon$.
Firstly, we develop an analytical solution to the linearized flow response to a Gaussian lift and drag force and use the results to establish a relationship between the local and far-field  velocity required to specify lift and drag forces. Then, focusing first on the lift force,  we find  $\epsilon$ and the force center location that minimize the square difference between the velocity fields induced by the Gaussian force and 
2D potential flow over Joukowski airfoils. 
We find that the optimal smoothing width $\epsilon^{\rm opt}$ is on the order of 14-25\% of the chord length of the blade, 
and the center of force is located at about 13-26\% downstream of the leading edge of the blade, for the cases considered.  
These optimal values do not depend on angle of attack and depend only weakly on the type of lifting surface.
To represent the drag force, the optimal width of the circular Gaussian drag force field is 
shown to be equal to the momentum thickness of the wake. 
It is then shown that an even more realistic velocity field can be 
induced by a 2D elliptical Gaussian lift force kernel, 
and the corresponding optimal semi-axes $\epsilon_x$ and $\epsilon_y$ lengths 
are determined using the velocity matching method. 
\end{abstract}


\maketitle

\footnotetext[2]{Please ensure that you use the most up to date
class file,
available from the WE Home Page at\\
\href{http://www3.interscience.wiley.com/journal/6276/home}{\texttt{http://www3.interscience.wiley.com/journal/6276/home}}}

\section{Introduction}
Numerical simulations of flow over wind turbines often cannot afford to resolve the entire rotating blade geometry and the associated flow details \citep{sorensen_aerodynamic_2011}.  Actuator disk models (ADM) greatly reduce resolution requirements by replacing the rotor by a distributed body force at the rotor disk location \citep{jimenez_advances_2007, calaf_large_2010}. However, the ADM misses important information about the instantaneous blade location and rotation, its detailed aerodynamic properties, etc. An intermediate approach is provided by  the  actuator line model (ALM) \citep{sorensen_numerical_2002}, a  technique for simulating lifting surfaces within flow simulations with characteristics that fall in between full blade resolution and the ADM. The  ALM has been particularly popular in large-eddy simulations (LES)  of wind turbine blades \citep{sorensen_numerical_2002, mikkelsen_analysis_2007, troldborg_numerical_2010, 
porte2011large, churchfield_numerical_2012, churchfield_large-eddy_2012, martinez-tossas_large_2014, jha2014guidelines, xie_self-similarity_2015}. 
At any blade cross-section, the local lift and drag forces, which are 
evaluated locally  using tabulated lift and drag coefficients as function of the local angle of attack, are imposed at points moving with the blades within the simulation domain.  
The actuator surface model (ASM) is another method to represent 
lifting surfaces \cite{shen2009actuator,Sibuet2010}.
In the ASM the blades are represented as a infinitesimal 
sheet with a vorticity source. Previous work has shown improvements in the flow fields
produced by the ASM compared to the ALM \cite{shen2009actuator}.
The present work focuses on the ALM because of its simplicity,
and its  wide and common usage in representing
wind turbine blades in  Large Eddy Simulations \cite{sorensen_numerical_2002}.

In the ALM, it is common practice to smooth the imposed force using a Gaussian kernel to distribute the imposed point force. The 3D kernel is given by
\begin{equation}
\eta_\epsilon = \frac{1}{\epsilon^3 \pi^{3/2}} e^{-r^2/\epsilon^2}.
\end{equation}
Here $r$ is the distance to the actuator point and
$\epsilon$ establishes the kernel width. This function
is introduced to prevent numerical issues that can arise from the application
of discrete body forces in a computational domain \citep{sorensen_numerical_2002}.
Other studies have addressed various aspects of this
kernel \citep{troldborg_numerical_2010}. Specifically, the effects of the
parameter $\epsilon$ on ALM predictions of  power and wake structures in wind turbines have been examined in previous work \cite{martinez-tossas_large_2014, jha2014guidelines}, where it was 
also hypothesized that $\epsilon$ should physically scale with the chord length of the blade, absent numerical constraints.
Although there are a number of guidelines on how to choose $\epsilon$ depending mostly on the numerics and grid resolution, 
no clear physical rationale for choosing a particular value of $\epsilon$ has been proposed to date.

In this work, we present a calculation of the optimal value of the Gaussian width $\epsilon$, and of the 
best location to center the force,
based on physical arguments instead of relying only on numerical justifications.  The optimal values we seek will be
based on the ability of the ALM to reproduce the induced velocity distributions as realistically as possible. 
An extension is presented for a 2D Gaussian kernel with different
smoothing lengths in the chord and thickness directions of the airfoil.

\section{Formulation}

We develop an analytical solution to the 2D flow generated when using the ALM with a smoothing length-scale $\epsilon$. Specifically, we consider the Euler equations in which a lift force (with circulation $\Gamma$) has been applied in the $y$-direction perpendicular to the free-stream velocity $U_{\infty} \ihat$ in the $x$-direction. We consider the flow in the frame of reference moving with the blade cross-section, so that in turbomachinery applications the far-field velocity $U_{\infty} \ihat$ includes the rotor tangential velocity and axial induction. In this frame, the corresponding solutions are denoted as the velocity and pressure fields ${\bf u}_\epsilon(x,y)=[u_\epsilon(x,y),v_\epsilon(x,y)]$ and $p_\epsilon(x,y)$ respectively. They are solutions to:
\begin{equation}
{\bf u}_\epsilon \cdot \nabla {\bf u}_\epsilon = - \nabla p_\epsilon/\rho - \jhat ~\frac{U_\infty \Gamma}{\epsilon^2 \pi}  ~e^{-({\bf x}-{\bf x}_0)^2/\epsilon^2}, ~~~~ \nabla \cdot {\bf u}_\epsilon=0, 
\label{eq:epsilon}
\end{equation}
subject to boundary condition ${\bf u}_\epsilon = U_{\infty} \ihat$ as $|{\bf x}| \to \infty$. Note that we are considering the flow in 2D sections and thus neglect 3D end effects. We are also neglecting Coriolis accelerations, and viscous and turbulence effects.  An analytical solution to the linearized problem, valid for small lift forces, will be derived in \S \ref{sec-solutionepsilon}, as a function of $\epsilon$.   To establish the best value of $\epsilon$, we seek to compare the obtained solution to the model problem with an ``exact'' solution of 2D flow past a lifting surface. As ground truth we use the potential flow solution to the Euler equations with the appropriate boundary conditions on the blade and the separation point at the trailing edge.  We denote the potential flow solution as the velocity and pressure fields, ${\bf u}_p(x,y)=[u_p(x,y),v_p(x,y)]$ and $p_p(x,y)$ respectively. These fields solve the Euler equations without the force, but with the appropriate boundary conditions at the blade surface:
\begin{equation}
{\bf u}_p \cdot \nabla {\bf u}_p = - \nabla p_p/\rho,~~~~ ~ \nabla \cdot {\bf u}_p=0, ~~~ {\bf u}_p \cdot {\bf n} = 0 ~~{\rm on~blade~surface},
\label{eq:potential}
\end{equation}
and, again, the boundary condition at infinity is ${\bf u}_p = U_{\infty} \ihat$ as $|{\bf x}| \to \infty$. The vector ${\bf n}$ is the unit vector normal to the blade surface. Also, the Kutta condition is implied with the separation point positioned at the trailing edge. For convenience, we use Joukowski airfoils and use the corresponding analytical solutions, recalled succinctly in \S \ref{sec-solutionpotential}.

Our goal  is to determine the values of $\epsilon$ and ${\bf x}_0$ for which the difference ${\bf u}_\epsilon(x,y) - {\bf u}_p(x,y)$ between the two velocity fields is minimized, in some sense. The precise definition of the associated error norm, its minimization, and results are described in \S \ref{sec-optimization}. It is important to recall that while the ideal velocity field is potential flow, the velocity field arising from Euler equation with a Gaussian body force is not potential since the body force is rotational.  Still, we expect that the position ${\bf x}_0$ and scale $\epsilon$ of the force can be chosen so as to best approximate the potential flow velocity field over a realistic lifting surface. 

Furthermore, we consider the case of a drag force, in which a  force in the $\ihat$ direction is applied with a Gaussian distribution of width $\epsilon_d$. In \S \ref{sec-drag} we use the analytical solution to the linearized problem to determine a correction to the local velocity to infer the far-field velocity. We discuss the possibility of applying the drag force using a kernel width $\epsilon_d$ that could differ from that for the lift force. 

\section{Linearized Euler equation with Gaussian body forces}
\label{sec-solutionepsilon}
 
Here, an analytical solution to Eq. \ref{eq:epsilon} is sought.
The equations are linearized about the free-stream velocity,
where the velocity perturbation $u^\prime << U_\infty$.
A similar equation with a drag force instead of a
lift force is used to understand the effects
of a local streamwise body force on the  flow. 
The analytical solutions are used to relate the 
local velocity sampled at the 
center of the Gaussian lift and drag force distributions with the far-field velocity needed in 
applications of the ALM. Then the solutions can be used to determine optimal kernel width. 

\subsection{Solution with Lift Force}
We begin by non-dimensionalizing the problem using $U_\infty$ as velocity scale, and the lifting surface's chord length $c$ as length scale. We denote non-dimensional variables using asterisks, i.e $x_*=x/c$, $y_*=y/c$, $\epsilon_*=\epsilon/c$, $u_*=u/U_\infty$, $v_*=v/U_\infty$, and dimensionless circulations $\Gamma_* = \Gamma/c U_\infty$ and 
$K_* = \Gamma_*/2\pi$. In order to eliminate pressure from equation \ref{eq:epsilon} we take its curl and obtain an equation for the scalar dimensionless vorticity $\omega_{\epsilon*}$ in the $z$-direction:
\begin{equation}
{\bf u}_{\epsilon *} \cdot \nabla \omega_{\epsilon *} = -
[ \nabla \times  \frac{2 K_*}{\epsilon_*^2}
 e^{-(x_*^2+y_*^2)/\epsilon_*^2} ~\jhat ~ ]_{z}.
\end{equation}
In expanded form, the equation reads
\begin{equation}
u_{\epsilon *} \frac{\partial \omega_{\epsilon *}}{\partial x_*}
+
v_{\epsilon *} \frac{\partial \omega_{\epsilon *}}{\partial y_*}
= 
4 x_*  \frac{K_*}{\epsilon_*^4} e^{-(x_*^2+y_*^2)/\epsilon_*^2}.
\end{equation}

We now consider the small $K_*$ behavior, and use a linear perturbation analysis around  $K_*=0$. The baseline solution
is uniform flow in the $x$-direction and zero vorticity, i.e.~$ {\bf u}_{\epsilon *} =  \ihat + {\bf u}_{\epsilon *}^\prime$ and $\omega_{\epsilon *} = \omega_{\epsilon *}^\prime
$ so that 
\begin{equation}
{\bf u}_{\epsilon *} \cdot \nabla  \omega_{\epsilon *} \approx  \frac{\partial \omega_{\epsilon *}^\prime}{\partial x_*} =
4 x_* \frac{K_*}{\epsilon_*^4} e^{-(x_*^2+y_*^2)/\epsilon_*^2}.
\end{equation}
Following  Ref.~\cite{james_r_forsythe_coupled_2015} 
(their Appendix) we integrate in $x_*$, and using the condition that the perturbation vorticity vanishes at infinity leads to 
\begin{equation}
	\omega_{\epsilon *}^\prime(x_*,y_*) = - \frac{2 K_*}{\epsilon_*^2} e^{-(x_*^2+y_*^2)/\epsilon_*^2}.
    \label{eq:vorticity}
\end{equation}
As shown  in \cite{james_r_forsythe_coupled_2015} for more general cases, the vorticity distribution is proportional to the local body force. 
This occurs because the source of vorticity is the curl of the body force, which when integrated along 
the streamlines yields a 
distribution proportional to the body force. 

The next step is to obtain the induced velocity from the vorticity distribution. 
The velocity field induced by a circular 
Gaussian vorticity distribution is a standard 
solution that can be obtained in polar coordinates, where
the circulation around a circle of radius $r_*$ centered at the force center 
is given by 
\begin{equation}
2\pi ~r_*  ~V_{\theta,\epsilon*}^\prime(r_*) = \int \int  \omega_{\epsilon *}^\prime dA_* = 
2\pi
\int_{0}^{r_{*\epsilon}}
 \frac{2 K_*}{\epsilon_*^2}  e^{-r_*'^2/\epsilon^2}~  r_*' dr_*'.
\end{equation}
Here $V_{\theta,\epsilon*}^\prime$ is the tangential component of the vorticity-induced perturbation velocity:
\begin{equation}  
V_{\theta,\epsilon*}^\prime(r_*)=  K_* ~\frac{1}{r_{*}}~ \left( 1-e^{-r_*^2/{\epsilon_*^2}} \right).
\end{equation}
The complete solution is the perturbation added to the base flow, which when expressed in Cartesian coordinates with the Gaussian kernel 
centered at ${\bf x}_{0*} = (x_{0*},y_{0*})$ reads 
\begin{equation}
u_{\epsilon*}= 1 +
K_*
\frac{y_* - y_{0*}}{(x_* - x_{0*})^2+(y_* - y_{0*})^2}
\left[1-e^{- \left( (x_* - x_{0*})^2+(y_* - y_{0*})^2 \right)/\epsilon_*^2} \right]
\label{eq:solutionueps}
\end{equation}
\begin{equation}
v_{\epsilon*}= - K_* \frac{x_* - x_{0*}}{(x_* - x_{0*})^2+(y_* - y_{0*})^2}
\left[ 1- e^{-\left( (x_* - x_{0*})^2+(y_* - y_{0*})^2 \right) / \epsilon_*^2} \right]
\label{eq:solutionveps}
\end{equation}

\subsection{Solution with Drag Force}
\label{sec-drag}

Often in applications of ALM, the distributed 
force also includes a drag component acting 
(in our coordinate system) 
in the $x-$direction. Let us consider the problem separately from the lift force, and 
consider the linearized problem when we apply a drag force $F_D$ 
(in 2D per unit length) using the ALM 
with a Gaussian kernel of width  $\epsilon_d$ 
(the subscript stands for ``drag''). 
The linearized evolution equation for the 
perturbation vorticity due to the imposed drag 
force now reads:  
\begin{equation}
U_\infty \ihat \cdot \nabla \omega^\prime_{\epsilon d} =
- \left[ \nabla  \times \left( \frac{F_D}{\rho \epsilon_d^2 \pi} e^{-(x^2+y^2)/\epsilon_d^2}~ \ihat \right)~ \right]_z.
\end{equation}
Normalizing and expanding yields 
\begin{equation}
\frac{\partial \omega^\prime_{\epsilon_d*}}{\partial x_*}  = 
-\frac{ c_d}{\epsilon_{d*}^4 \pi}~ y_* ~e^{-(x_*^2+y_*^2)/\epsilon_{d*}^2},
\end{equation}
where $c_d = F_D/(\frac{1}{2} \rho U_\infty^2 c)$ is the drag coefficient. Integration in $x$ between $-\infty$ to $x_*$ leads to
a vorticity distribution given by 
\begin{equation}
\label{eq:omegadrag}
\omega^\prime_{\epsilon_d*}(x_*,y_*)   =   
-\frac{c_d}{ 2 \sqrt{\pi} \, \epsilon_{d*}^3} 
\, y_* \, e^{-y_*^2/\epsilon_{d*}^2}
\left[1+{\rm erf}(x_*/\epsilon_{d*}) \right].
\end{equation}
The x-direction velocity in the far-field (at $x_* >>1$ where the vertical velocity  $v_{\epsilon*}$ is negligible) can be obtained by integrating $\partial u_{\epsilon_d*}/\partial y_* = - \omega^\prime_{\epsilon_d*}$, leading to the defect perturbation velocity distribution 
\begin{equation}
u^\prime_{\epsilon_d*}(x_*,y_*)  =   - \frac{c_d}{2\epsilon_{d*} \sqrt{\pi}} ~e^{-y_*^2/\epsilon_{d*}^2} ~~~~~\mbox{for}~~~  x_* >>1.
\label{eq:uprimewake}
\end{equation}
 The `initial' wake profile 
imposed by the Gaussian body force in the ALM 
in the spanwise $y_*$ direction is therefore 
a Gaussian of width $\epsilon_{d*}$. 
Its deficit magnitude grows initially in the 
$x$-direction within the kernel region.

\subsection{Velocity Sampling in  ALM}
\label{sec-velsampling}

 In ALM one commonly samples the velocity at the center of the applied 
Gaussian force and uses this velocity instead of $U_{\infty}$ to determine forces based on given lift and drag coefficients
 \citep{martinez-tossas_large_2014}. 
This is usually done in the center of the Gaussian kernel 
\cite{sorensen_numerical_2002,troldborg_numerical_2010,martinez-tossas_large_2014},
but others have tried different approaches \cite{james_r_forsythe_coupled_2015}.
Here we use the analytical solutions obtained above to clarify where the velocity may be sampled and how to correct the sampled velocity. 
We consider the effects of lift and drag separately. 

In terms of lift,  we immediately note from the analytical solution (Eq. \ref{eq:solutionueps})  that ${\bf u}_{\epsilon *}(x_0,y_0)=U_\infty \ihat$, 
i.e.~at the center point of the imposed force (also the vortex center) the perturbation velocity vanishes. 
Thus the resulting velocity there equals the free-stream 
velocity, even in the presence of an 
imposed ALM lift force.  Therefore, sampling the velocity at the center of the Gaussian applied lift force provides the free stream velocity automatically 
thus  justifying the more common approach to sample velocities in ALM applications  
(see \cite{james_r_forsythe_coupled_2015}) for more general treatments).

In the presence of  drag, however,  the situation is different as the velocity at the force center is affected by the imposed force.  	
In order to quantify this effect, we must find the perturbation velocity at the center of the Gaussian drag force that results from the perturbation vorticity distribution 
in Eq. \ref{eq:omegadrag}. 
To this effect we use the 2D Biot-Savart equation evaluated at the origin (center of Gaussian):
	\begin{equation}
	u^\prime_{\epsilon_d*}(0,0) = \frac{1}{2 \pi}
	\int_{-\infty}^{\infty}
	\int_{-\infty}^{\infty}
	\frac{y_* \, \omega^\prime_{\epsilon_d*}(x_*, y_*)}
	{\left(x_*^2 + y_*^2\right)}
	\, dx_* \, dy_*
	\label{eq:biot}
	\end{equation}
	A change of variables is used
	to express $x$ and $y$ in terms of $\epsilon_d$
	by $x_{*\epsilon} = x_* / \epsilon_{d*}$.
	Replacing the vorticity distribution of Eq. \ref{eq:omegadrag}, the perturbation velocity at the origin can then be expressed according to
	\begin{equation}
	\frac{u^\prime_{\epsilon_d*}(0,0) \, \epsilon_{d*}}
	{c_d} = - \frac{1}{4 \pi^{3/2}}
	\int_{-\infty}^{\infty}
	\int_{-\infty}^{\infty}
	\frac{y_{*\epsilon}^2}
	{\left(x_{*\epsilon}^2 + y_{*\epsilon}^2\right)} \,
	e^{-y_{*\epsilon}^2} \left[ 1 + \erf (x_{*\epsilon}) \right]
	\, dx_{*\epsilon} \, dy_{*\epsilon}
	\label{eq:dragUintegral}
	\end{equation}
	By anti-symmetry of the error function, the $x_{*\epsilon}$ integration only contains the 
	$1/(x_{\epsilon*}^2 + y_{\epsilon *}^2)$ term and thus
	\begin{equation}
	- \frac{u^\prime_{\epsilon_d*} \, \epsilon_{d*}}
	{c_d} = \frac{1}{4\sqrt{\pi}}
	\int_{-\infty}^{\infty}
	\vert y_{*\epsilon} \vert \,e^{-y_{*\epsilon}^2} 
	dy_{*\epsilon}
	= \frac{1}{4 \sqrt{\pi}} \approx 0.141
	\label{eq:dragUintegral2}
	\end{equation}	

From Eq.  \ref{eq:dragUintegral2} the following observation can be made.
The streamwise velocity at the center due to a drag body force is given by 
\begin{equation}
u_{\epsilon_d}(x_0, y_0) = U_\infty + u_{\epsilon_d}^\prime(x_0, y_0) = U_\infty\left(1- \frac{1}{4 \sqrt{\pi}} \, c_d \, \frac{c}{\epsilon_d}\right).
\end{equation}
Therefore, based on $u_{\epsilon_d}(x_0, y_0)$, the velocity sampled at the center of the Gaussian, the free stream reference velocity may be estimated according to
\begin{equation}
  U_\infty   = \frac{u_{\epsilon_d}(x_0, y_0)}{1- \frac{1}{4 \sqrt{\pi}} \, c_d \, \frac{c}{\epsilon_d}},
  \label{eq:correction}
\end{equation}
which can then be used in the determination of lift and drag forces. 

The degree of nonlinearity is given by the ratio $u^\prime/U_\infty$, i.e. $\frac{1}{4 \sqrt{\pi}} \, c_d \, \frac{c}{\epsilon_d}$. For 
$c_d<1$, even choosing $\epsilon_d \sim c$ we see that the nonlinearity is less that 0.14. We have simulated a Gaussian force in a fully nonlinear 2D Navier-Stokes  solver
(not shown) and verified empirically that 
the correction factor in Eq. \ref{eq:correction} 
is accurate at least up to 
$\frac{1}{4 \sqrt{\pi}} \, c_d \, 
\frac{c}{\epsilon_d} \approx 0.28$
so that we believe the correction 
in Equation \ref{eq:correction}
can be applied quite broadly in practice. 

\section{Potential flow over Joukowski airfoil}
\label{sec-solutionpotential}
Next, we briefly review potential flow solution for flow over a 
Joukowski airfoil to be used as ``ground truth'' to compare the velocity 
field induced by a Gaussian lift force distribution. 
Potential flow over a Joukowski airfoil is found by mapping the solution of flow over a lifting cylinder
onto the new complex space. It involves the complex velocity $w(\zeta)$ 
where $\zeta$ is the complex position variable $\zeta=\chi + i\psi$.
Again $U_\infty$ is the far-field velocity, and now $R$ is the cylinder radius,  $\alpha$ is the angle of attack,  
$\mu$ is a shift in the complex plane, and $\Gamma = 2 \pi K$ is the circulation.  
Symmetric and cambered airfoils can be generated by shifting the solution
in the complex plane by $\mu$.
Again the equations are made non-dimensional using the chord length for the resulting airfoil $c$: $\zeta_* = \zeta/c$,  
and the dimensionless circulation  $K_* =  {K}/{U_\infty c}$ is the same as that implied by the 
Gaussian body force considered in the prior section.  
The complex velocity \citep{katz_low-speed_2001} can now be written in non-dimensional form according to
\begin{equation}
  w_*(\zeta_*) = \zeta_* e^{-i\alpha} + \frac{R_*^2}{\zeta_*-\mu_*} e^{i\alpha}
  + i K_* \log \left( \zeta_* -\mu_* \right).
\end{equation}
The Joukowski Transform is defined by
${z}_*(\zeta_*) e^{i\alpha} = \zeta_* +  {l_*^2}/{\zeta_*}$,
where $l_*$ is the length chosen such that the intersect
in the real axis of the circle becomes the trailing edge in the
transformed coordinate system. The transformed plane coordinate $z_*$ is given by $z_*=x_* + i y_*$.
The inverse transform is
\begin{equation}
  \zeta_*({z}_*) = \frac{1}{2} {z}_*e^{i\alpha} + \frac{{z}_*e^{i\alpha}}{2} 
  \left(  1 - \frac{4 l_*^2}{{z}_*^2e^{2i\alpha}}\right)^{1/2}
\end{equation}
The velocity in the transformed ${z}_*$ plane is given by:
\begin{equation}
  u_*({z}_*) - i v_*({z}_*) = \frac{dw}{d{z}_*} = \frac{dw}{d\zeta_*} \frac{d\zeta_*}{d{z}_*}
\end{equation}
Finally then, for a Joukowski airfoil, the potential flow  velocity components are given by

\begin{equation}
 u_{p*}(x_*,y_*) = u_{p*}({z}_*)  =  {\Real} \left[ \frac{ e^{-i\alpha} - 
   {e^{i\alpha}}
   { (\zeta_*({z}_*)-\mu_*) 
   ^{-2}} 
   + 
   i K_*  
   /(\zeta_*({z}_*)-\mu_*)
   \ 
   }
   {\left( 1 -
   l_*^2 / (\zeta_*({z}_*)-\mu_*)
   ^{2} \right) e^{-i\alpha}}
   \right] ,
   \label{eq:upJouk}
\end{equation}

\begin{equation}
  v_{p*}(x_*,y_*) = v_{p*}({z}_*)  =  - {\Imag} \left[ \frac{ e^{-i\alpha} - 
   {e^{i\alpha}}
   { (\zeta_*({z}_*)-\mu_*) 
   ^{-2}} 
   + 
   i K_*  
   /(\zeta_*({z}_*)-\mu_*)
   \ 
   }
   {\left( 1 -
   l_*^2 / (\zeta_*({z}_*)-\mu_*)
   ^{2} \right) e^{-i\alpha}} 
   \right]
   \label{eq:vpJouk}
\end{equation}
The value of $K_*$ is chosen according to the Kutta condition
to have the flow leave the trailing edge smoothly.
For a symmetric airfoil this is expressed as
$
  K_* = 2  R_* \sin(\alpha),
$
where $R_*$ is the complex plane cylinder radius.
We consider several cases of Joukowski airfoil: flat plate, a symmetric airfoil and cambered airfoil. 
In all cases the blade has chord length $c$ and is at an angle of attack $\alpha$, which selects the value of $K_*$. This is also the value used to reproduce the lift force in the Gaussian kernel solution derived in \S \ref{sec-solutionepsilon}.

\section{Minimizing the difference between two solutions}
\label{sec-optimization}

\begin{figure}
\centering
\includegraphics[width=0.3\linewidth]
{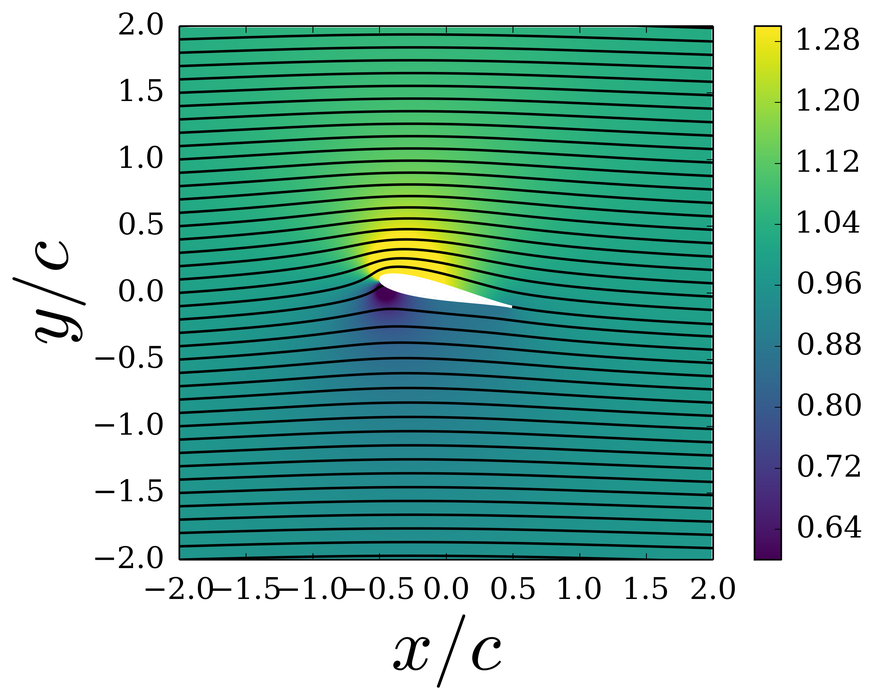}~
\includegraphics[width=0.3\linewidth]
{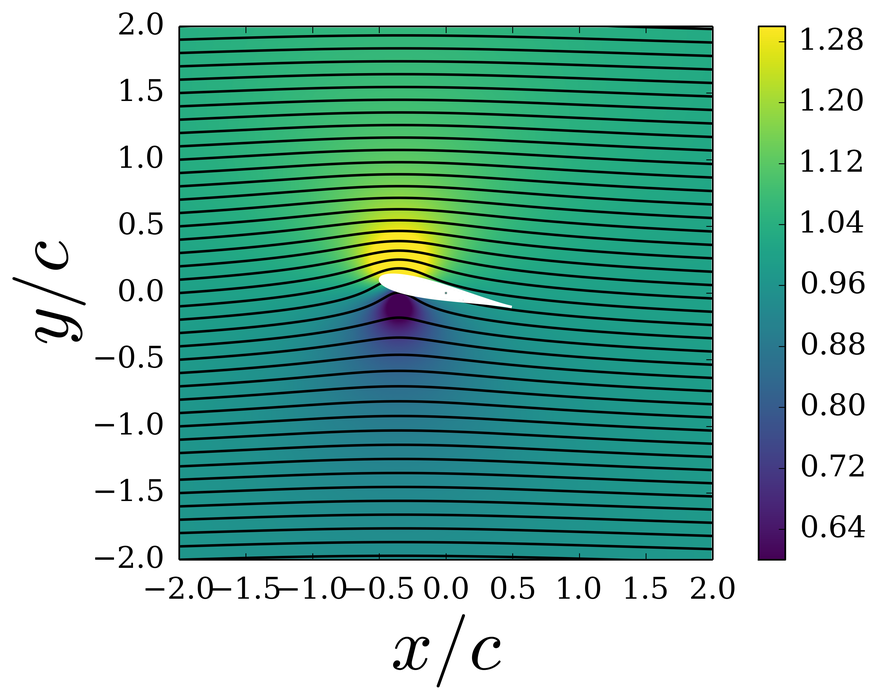}~
\includegraphics[width=0.3\linewidth]
{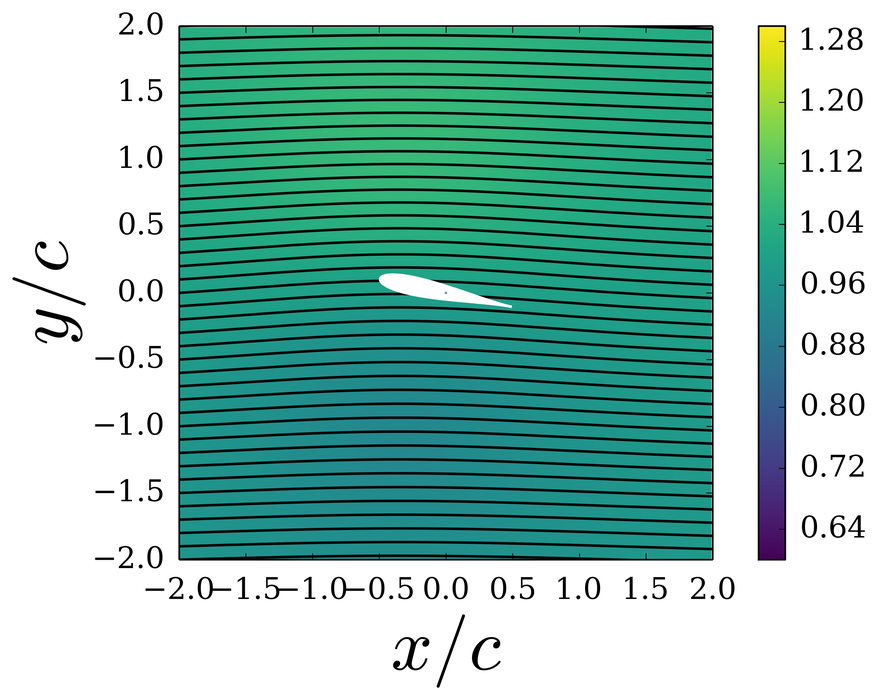}
\caption{Contour plots of velocity magnitude and streamlines. Left: Potential flow solution for a symmetric Joukowski airfoil at $\alpha=12^o$. Middle: 
Solution to linearized Euler equation with Gaussian body force with the optimal value of $\epsilon_*$. Right: 
Solution to linearized Euler equation with Gaussian body force with $\epsilon_*=1$.}
\label{fig:epsilonContours}
\end{figure}

The solutions for flow over a Joukowski airfoil and flow over a  Gaussian 
distributed lift force have been obtained. 
The latter depends upon the width parameter $\epsilon_*$ and position ${\bf x}_{0*}$ of the Gaussian kernel. 
To simplify, we assume the location of the force is located on the chord of the lifting surface and thus its position is parameterized by a single
chord location $s_{0*} = s_0/c$. 
The dimensionless chord location $s_*$ is defined such that  $s_* = -1/2$ at the leading edge and $s_*=+1/2$ at the trailing edge.  
Projecting the force center position onto the Cartesian coordinate system the 
location of the points becomes $x_{0*} = s_{0*} \cos(\alpha)$ and 
$y_{0*} = - s_{0*} \sin(\alpha)$.
Figure \ref{fig:epsilonContours} shows velocity magnitude contours with
streamlines for a symmetric airfoil with $\alpha=12^o$.
The potential flow solution is on the left, with the solution to the linearized Euler equation with body force using the optimum value
of $\epsilon_*$ (to be determined as explained below) 
shown in the middle panel, and the solution with $\epsilon_*=1$  \citep*{martinez-tossas_large_2014} to the right.
As can be seen, the velocity distribution generated by a Gaussian body force  in the vicinity of the blade  depends rather strongly on $\epsilon_*$, even though
the integrated quantities such as lift, momentum and very far-field perturbation velocity are, by definition, the same in all cases.
This provides a motivation to establish, in a quantitative manner, the optimum value of $\epsilon_*$ and its position $s_{0*}$.

The optimal values of the parameters $\epsilon_*$ and $s_{0*}$ are found by considering the following L2 norm of the difference between the two solutions:  
\begin{equation}
E_u^2 (\epsilon_*, s_{0*}) = \frac{1}{A_*} \int \int 
\left[
( u_{\epsilon*}  - u_{p*}  )^2 +
( v_{\epsilon*}  - v_{p*} )^2 
\right] f[\zeta_*(x_*,y_*)]~ dx_*dy_*
\label{eq:err}
\end{equation}
where $f(\zeta_*)$ is a mask to establish when the solution is outside of 
the airfoil area (i.e.~$f(\zeta_*)=0$ when $\zeta_*$ is inside the circle in the $\zeta_*$ plane, and one otherwise),
and $A_*=A/c^2$ is a fixed reference area, chosen to be
the square of the chord, i.e.~$A_*=1$ (and $A=c^2$).

\begin{figure}
\centering
\includegraphics[width=0.49\linewidth]{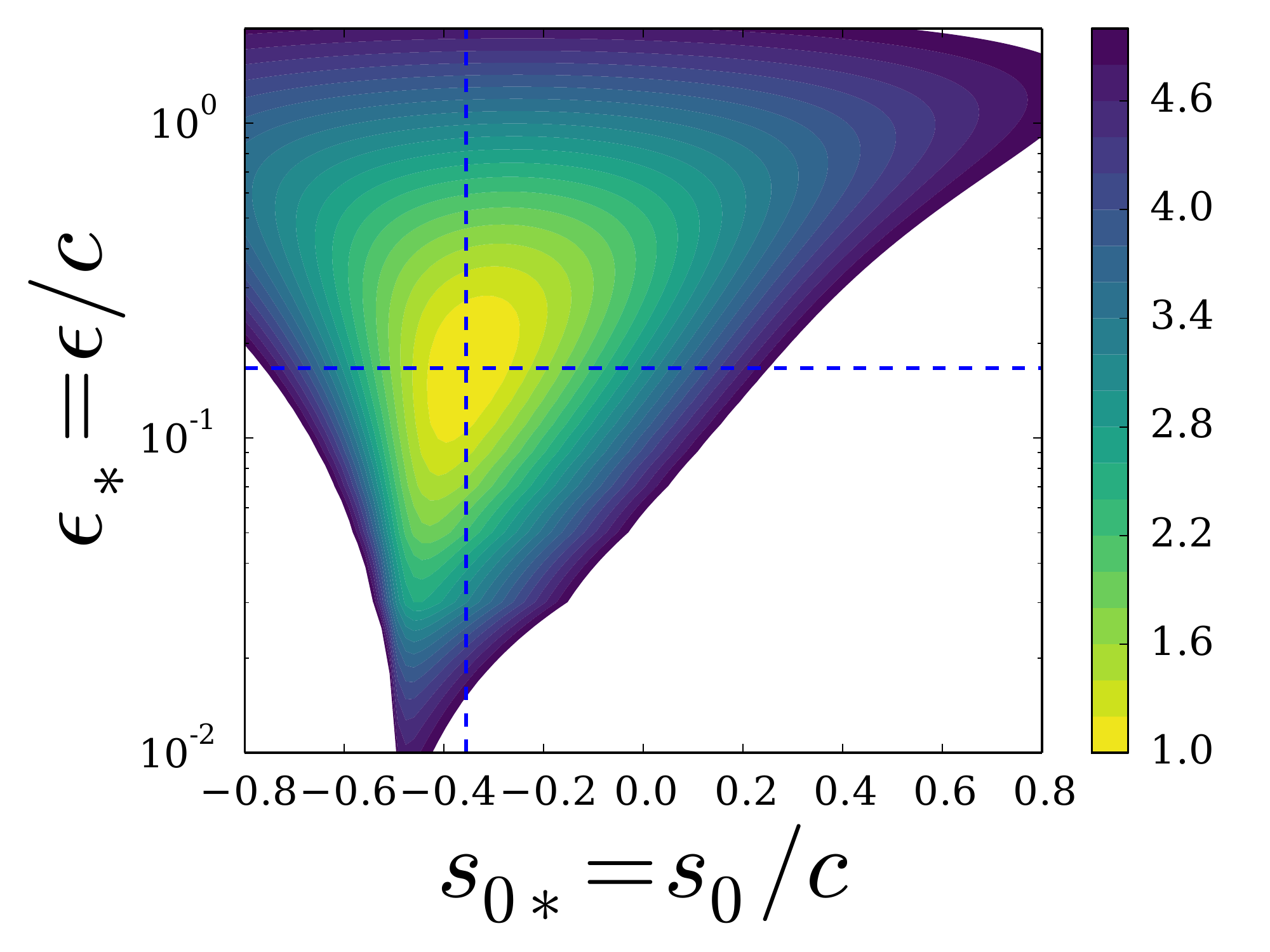}
\includegraphics[width=0.49\linewidth]{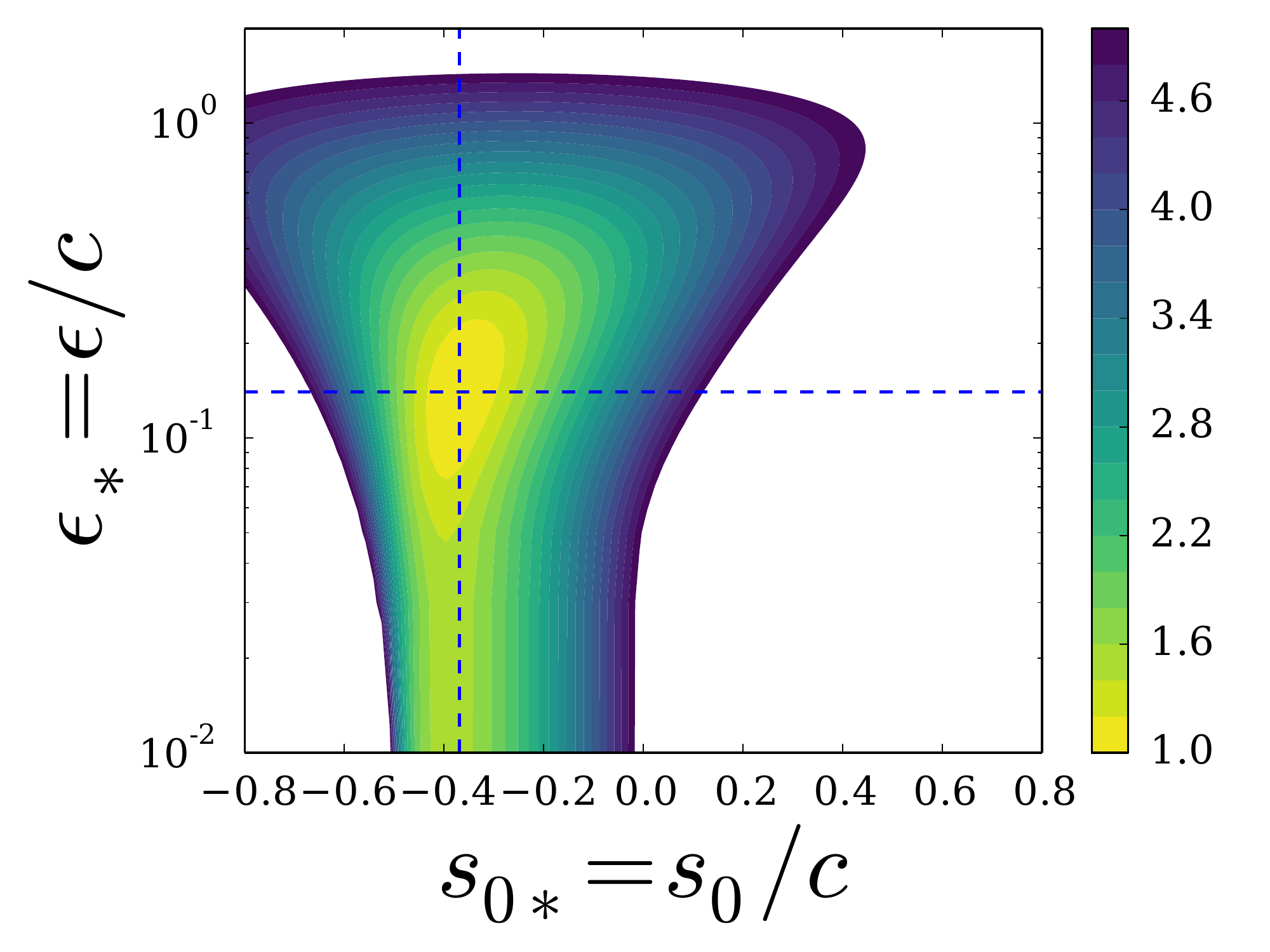}
\caption{Square error ($E_u^2/E_{u-min}^2$) contours as function of force center position $s_0/c$ and Gaussian width $\epsilon_*=\epsilon/c$ for the case with angle of attack $\alpha=12^o$. Results are shown for a flat plate (left) and for a symmetric Joukowski airfoil (right).
Vertical and horizontal lines mark the optimal value of the chord position $s_{0*}^{\rm opt}$ and $\epsilon_{*}^{\rm opt}$ respectively.}
\label{fig:contourError}
\end{figure}

The error is evaluated for the flat plate and symmetric Joukowski airfoil for a number of angles of attack (and implied circulations $K_*$). 
For each case, the error is evaluated by performing the integration in equation \ref{eq:err} for a range of
$(\epsilon_*, s_{0*})$ values. Results are shown in Figure \ref{fig:contourError} with dashed lines
showing the optimum values $(s_{0*}^{\rm opt},\epsilon_{*}^{\rm opt})$. 
The error is normalized with the square error at the minimum point. 
The minimum at $(s_{0*}^{\rm opt},\epsilon_{*}^{\rm opt})$  is found using 
the L-BFGS-B local minimizer \citep{byrd_limited_1995, oliphant_python_2007}.

\begin{figure}
\centering
\includegraphics[width=0.49\linewidth]{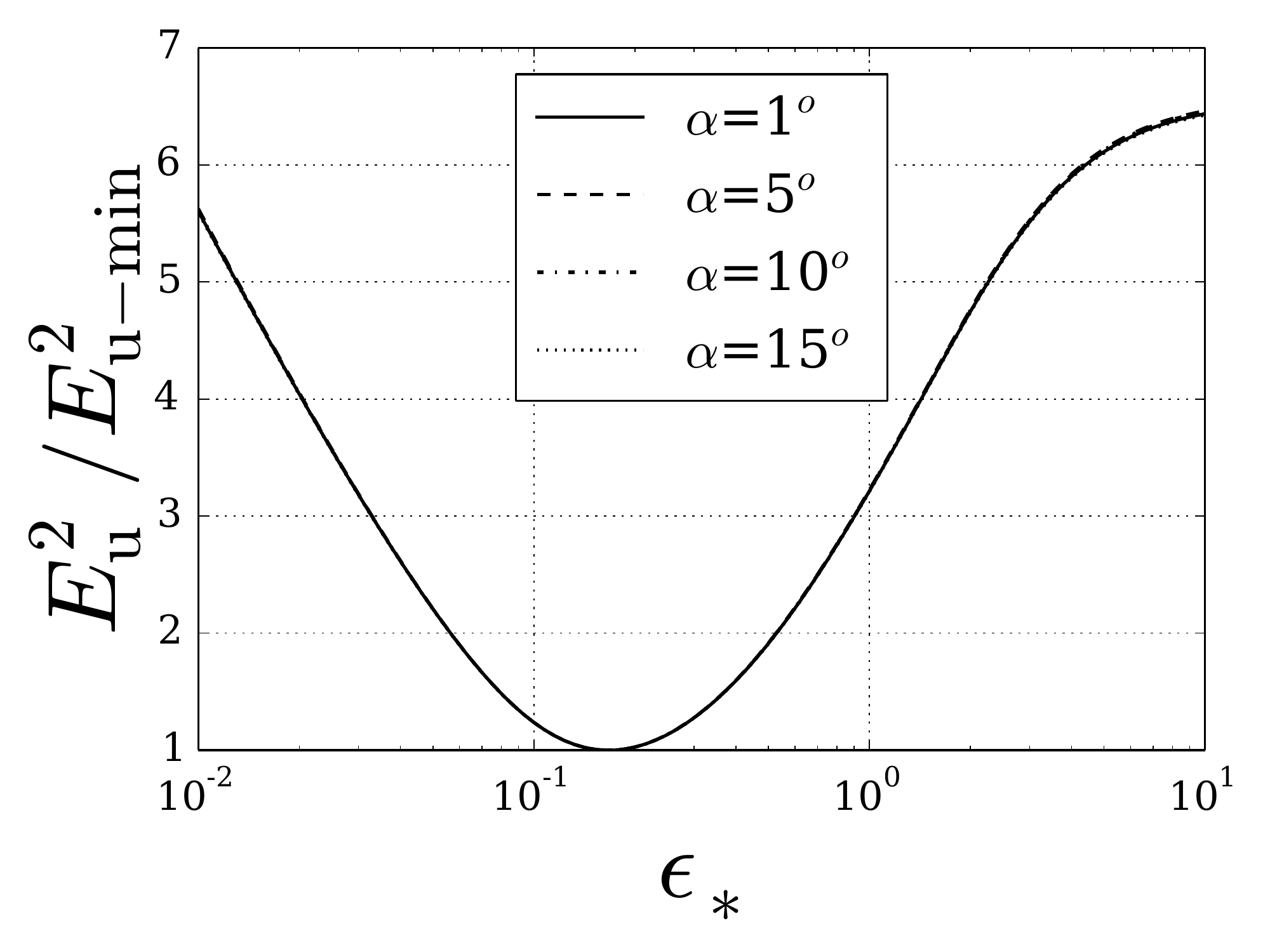}
\includegraphics[width=0.49\linewidth]{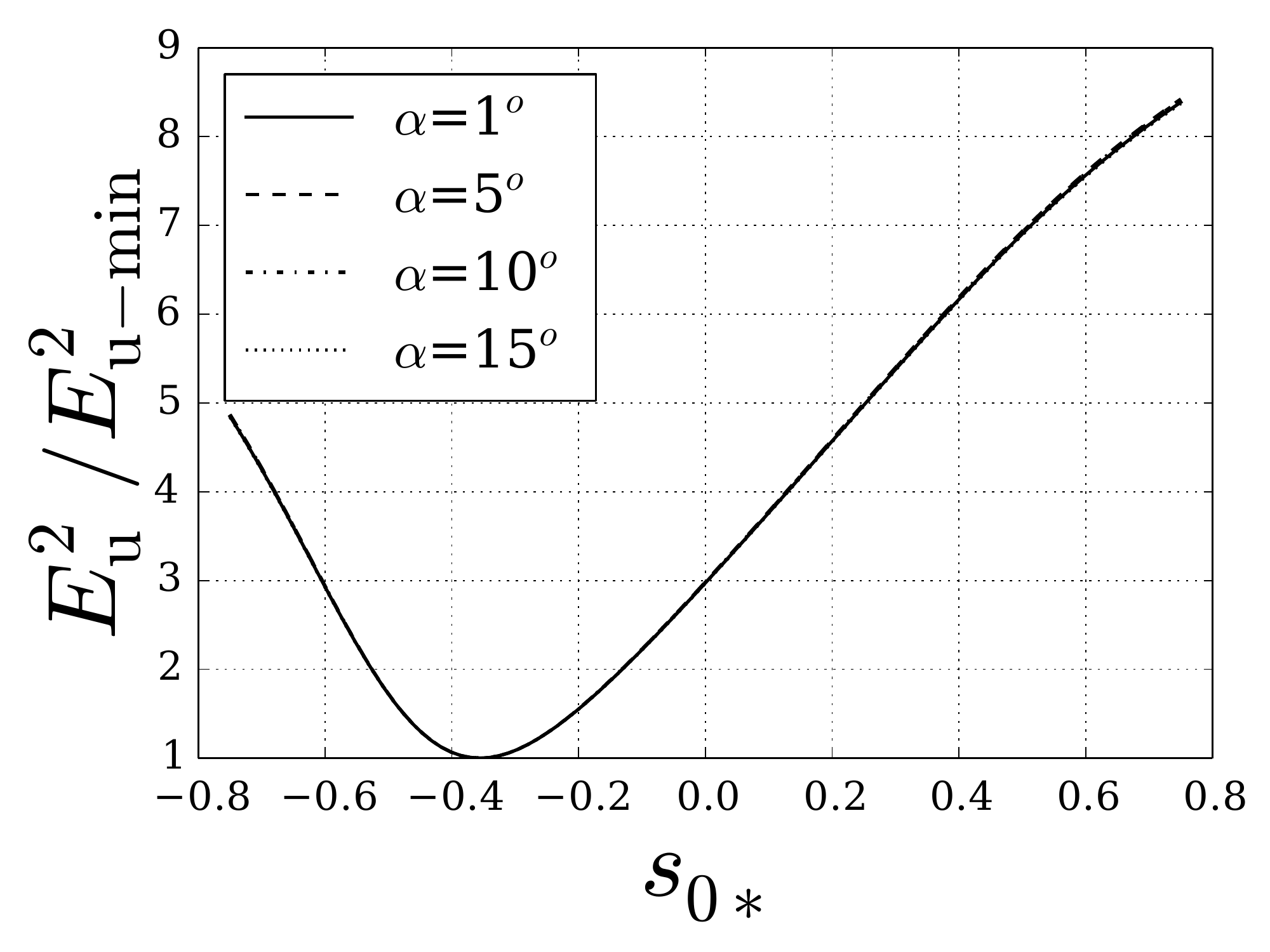}
\caption{Normalized square error for the flat plate as a function 
of $\epsilon_*$ (at $s_{*0}=-0.36$) and 
the chord position $s_{*0}$ (at $\epsilon_*=0.17$ for different angles of attack.}
\label{fig:epsilonlog1DError}
\end{figure}

\begin{figure}
\centering
\includegraphics[width=0.49\linewidth]{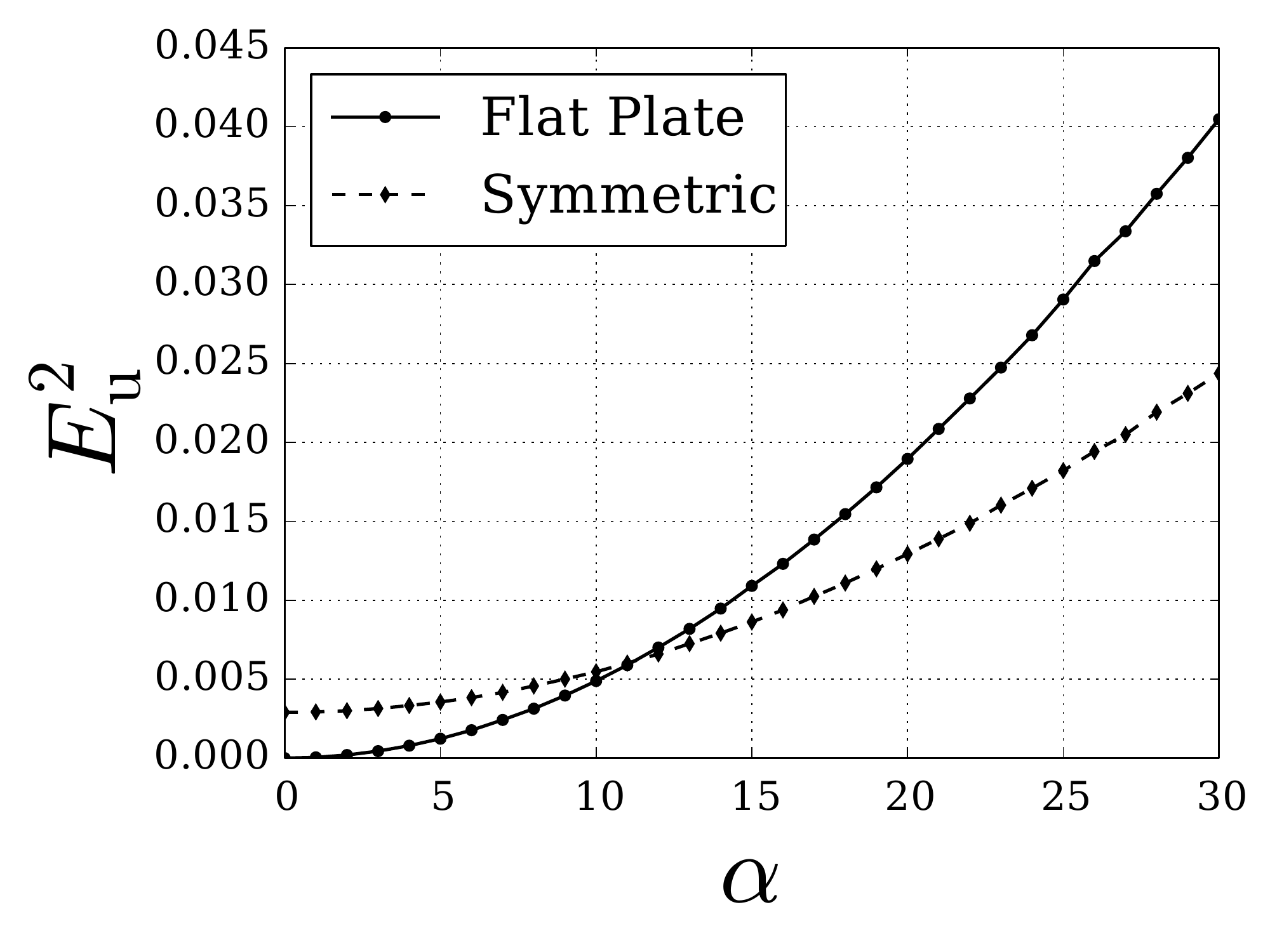}
\caption{Squared error at the optimum values $\epsilon^{\rm opt}_*$ 
and chord position $s_{*0}^{\rm opt}$ as a function of angle of attack $\alpha$.}
\label{fig:aoaError}
\end{figure}

The optimal values so obtained are listed in Table \ref{tab:optimum}. 
\begin{table}
  \begin{center}
\def~{\hphantom{0}}
  \begin{tabular}{ccc}
    Airfoil &  Optimum width $\epsilon^{\rm opt}/c$ & 
     Optimum position $s^{\rm opt}_0/c$ \\ 
    \hline  Flat Plate & 0.17  &  -0.36  \\
    Symmetric & 0.14 to 0.17  &  -0.37 to -0.35  \\
    Cambered  & 0.14 to 0.25   &  -0.37 to -0.24 \\
  \end{tabular}
  \caption{Optimum width and position of Gaussian force  for different airfoils.}
  \label{tab:optimum}
  \end{center}
\end{table}
The analysis is repeated for various angles of attack. We obtain essentially the same optimal values, independent of angle of attack  $\alpha$. This can be seen clearly in Figure 
\ref{fig:epsilonlog1DError} where the normalized error surface for the flat plate case is shown along the optimal values for several angles of attack. 
The curves show excellent collapse for different angles of attack, which indicates
that the optimum values of $\epsilon_*$ and chord position $s_{0*}$
are independent of $\alpha$.  
While the normalized error is independent of angle of attack,
the magnitude of the error depends on $\alpha$.
Figure \ref{fig:aoaError} shows the error for a 
flat plate and the symmetric airfoil as function of angle of attack.
The magnitude of the non-dimensional  error for angles of attack
from 0$^o$ to 15$^o$ ranges between 0 and 0.01 which translates to an average error of about 0 to 10\% (of $U_\infty$) in the 
perturbation velocity field in the relevant region. 

\begin{figure}
\centering
\includegraphics[width=0.49\linewidth]{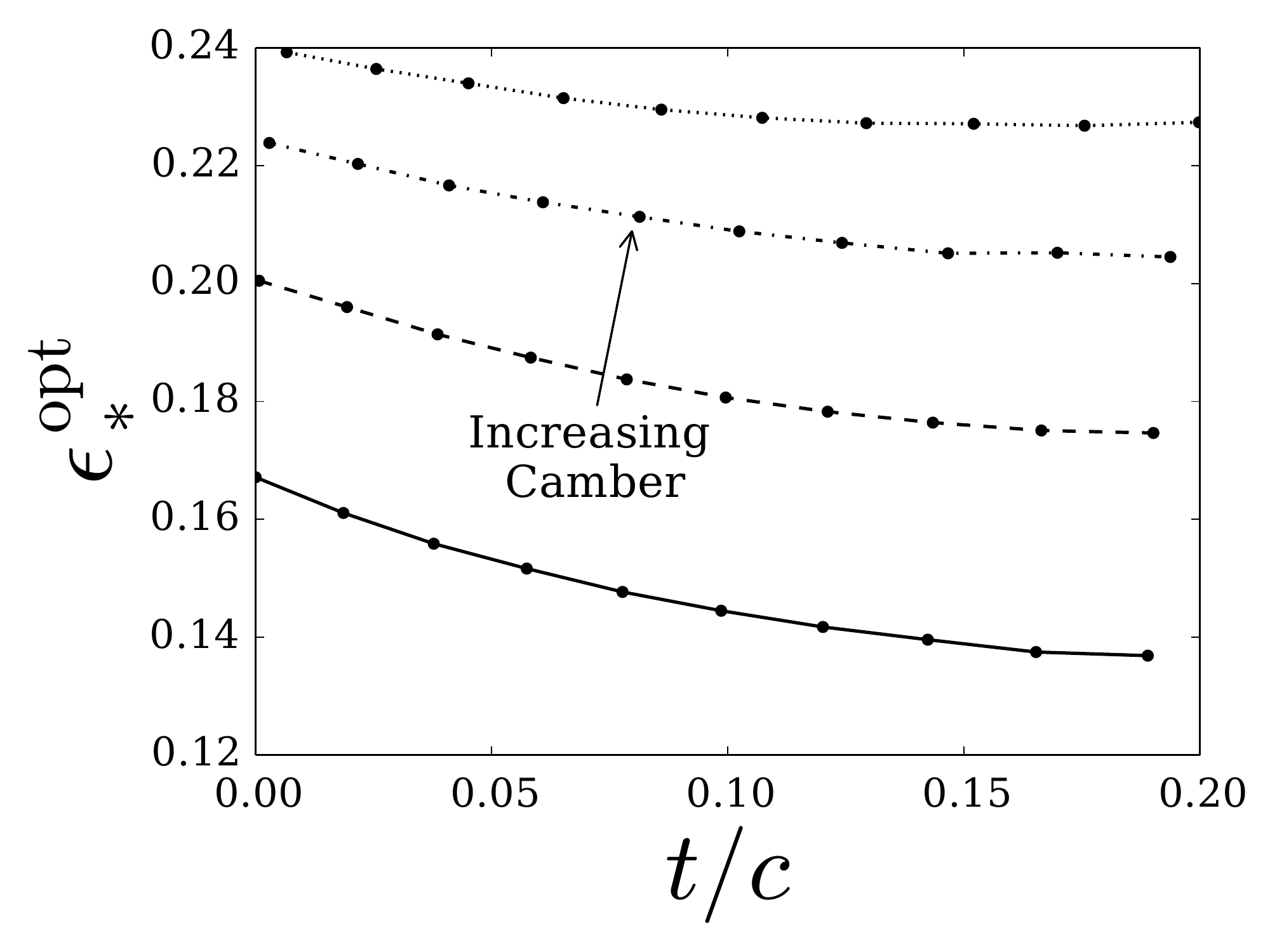}
\includegraphics[width=0.49\linewidth]{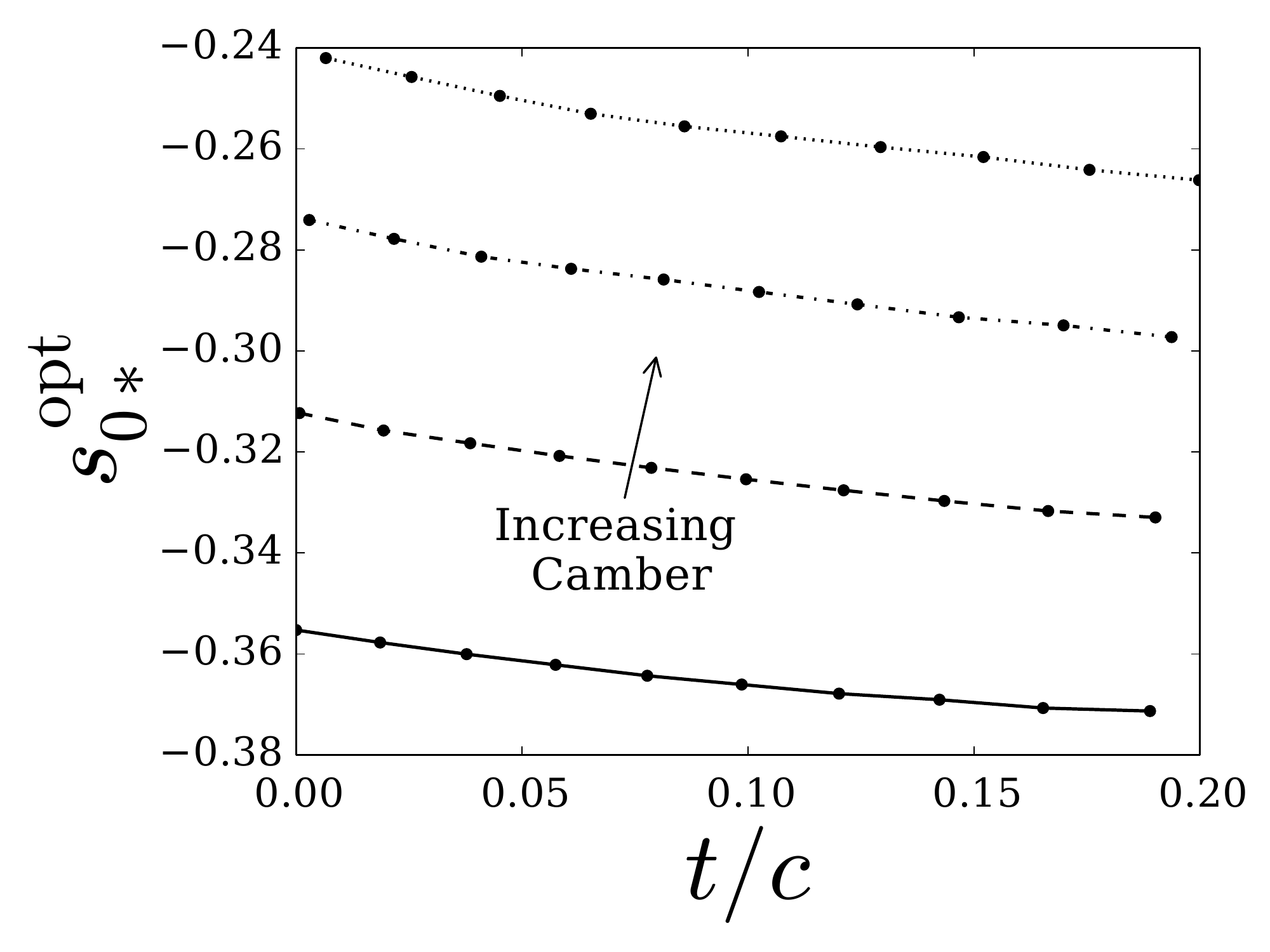}
\caption{Optimum values $\epsilon^{\rm opt}_*$ 
and chord position $s_{*0}^{\rm opt}$ as function of thickness $t/c$
for different camber with angle of attack $\alpha=12^o$.}
\label{fig:camberLines}
\end{figure}

The optimum values are a function of the thickness and camber of the airfoil.
Figure \ref{fig:camberLines} shows this dependency for a range of camber values and
airfoil thicknesses.
Camber and thickness are included by shifting the potential flow solution in the imaginary plane
$\zeta_*$ by a range from $\mu / R = 0$ (flat plate) to $\mu / R = -0.1 + 0.1i$ (cambered airfoil).
This range is chosen to match typical representative airfoil values \citep{katz_low-speed_2001}.
The solid line in Figure \ref{fig:camberLines}
represents the case for a symmetric airfoil.
Larger cambered airfoils require larger values of $\epsilon_*$
and the chord position of the Gaussian $s_{*0}^{\rm opt}$ is
closer to the quarter chord.
This placement allows for the streamlines to deform
following the cambered airfoil surface.
With more camber, the streamlines deform in a smoother
way following the airfoil surface, as opposed
to having a sharper deformation near the leading edge.
As the airfoil becomes thicker, $\epsilon^{\rm opt}_*$ 
becomes smaller.
The smaller $\epsilon^{\rm opt}_*$ allows the 
streamlines to deform more and align more closely
with the airfoil surface.

\begin{figure}
\centering
\includegraphics[width=0.49\linewidth]{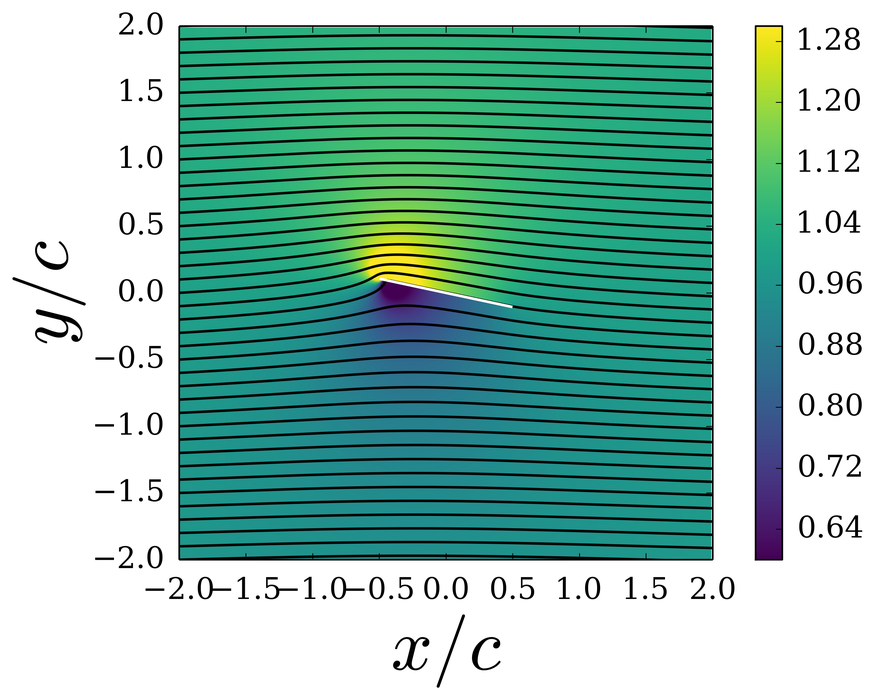}
\includegraphics[width=0.49\linewidth]{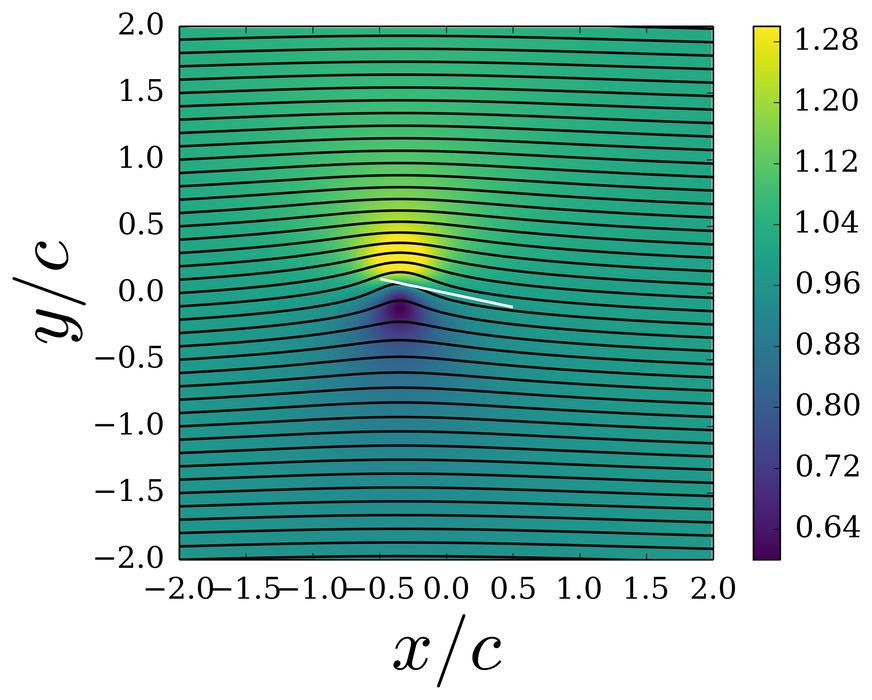}
\caption{Visualization of potential flow solution over a Joukowski airfoil (left) and body force solution (right)
for a flat plate. Colors  indicate the magnitude of the velocity field while solid lines indicate streamlines.}
\label{fig:flatPlate}
\end{figure}

\begin{figure}
\centering
\includegraphics[width=0.49\linewidth]{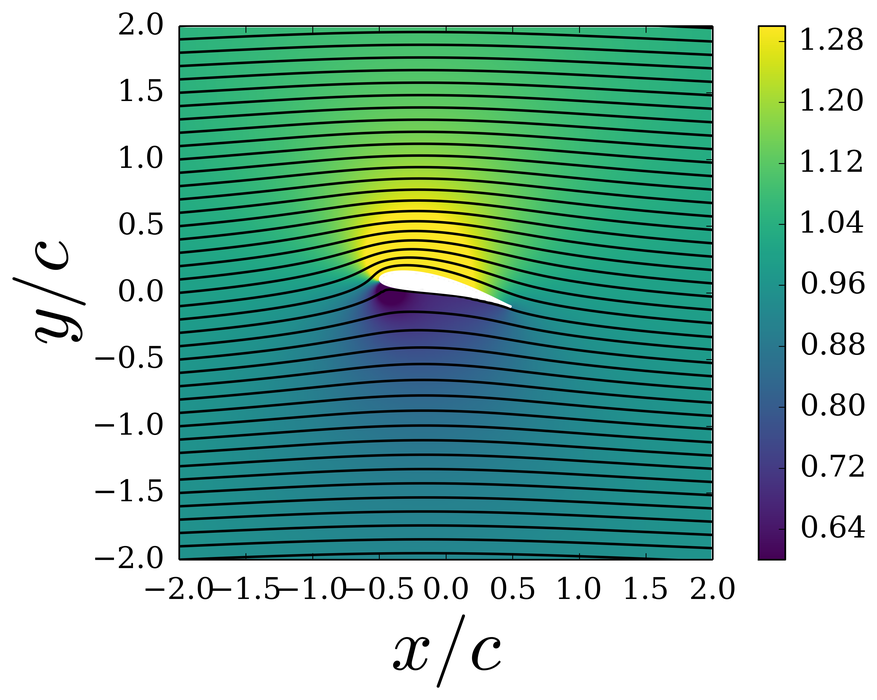}
\includegraphics[width=0.49\linewidth]{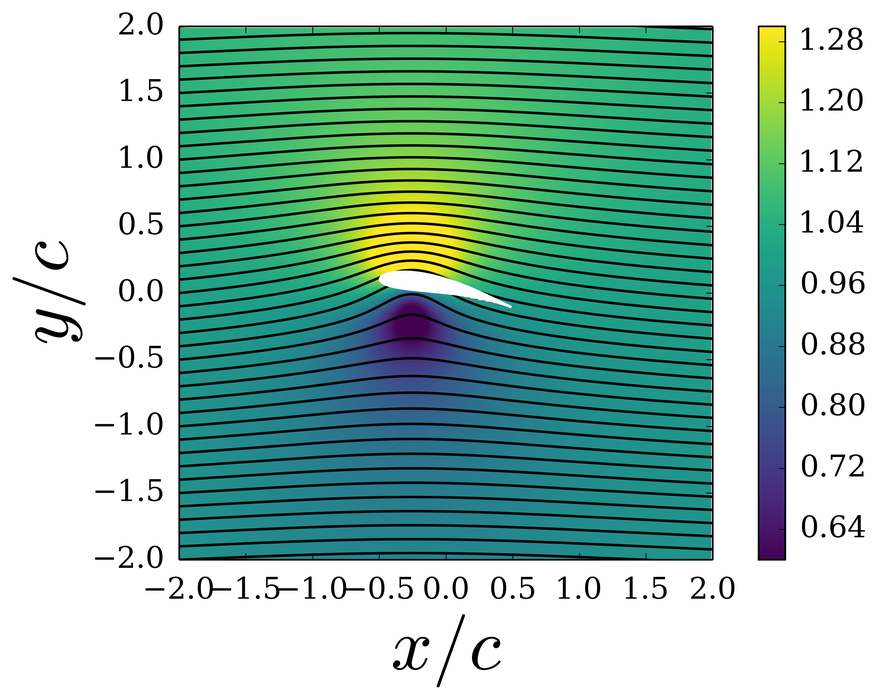}
\caption{Same as figure \ref{fig:flatPlate} but for cambered airfoil case.}
\label{fig:Cambered}
\end{figure}

Use of  $\epsilon^{\rm opt}/c$ and $s^{\rm opt}_{0}/c$ gives a flow field induced by the Gaussian force field 
which is as close as possible to the potential flow solution. In order to visualize the two velocity fields, in 
Figures \ref{fig:flatPlate}
and \ref{fig:Cambered} we compare streamlines and velocity magnitudes for the Joukowski potential flow solution and the model 
Gaussian force induced velocity field, for the cases of a flat plate and a cambered airfoil at $\alpha=12^o$. The case of symmetric airfoil has already been shown in 
Figure \ref{fig:epsilonContours} (left and middle panels). 
The solutions are qualitatively similar with the main features of
the flow being reproduced. Differences still exist as the streamlines in the potential flow solution are deformed when approaching the leading edge
and leaving the trailing edge smoothly obeying the boundary condition at the blade surface. Conversely, in the body force solutions the streamlines can pass through the airfoil, as expected from a solution without physical boundaries on the airfoil.

It is worthwhile noticing that while $\epsilon^{\rm opt} \sim \textit{O}(0.2 c)$ represents a fairly small fraction of the chord, the induced velocity field, being the integral of the Gaussian vorticity distribution, has a footprint at a scale that extends to a scale on the order of $c$. The optimization being based on the differences among the two velocity fields emphasizes the region near the blade at distances on the order of $c$. It is also noted that in the far field both solutions agree exactly with the ideal vortex tangential velocity distribution that decays as $\Gamma / (2\pi r)$, because by construction both the Gaussian forced velocity and the potential flow solution share the same imposed circulation and free-stream velocity.

\subsection{Optimal drag kernel width}

Returning now to the issue of representing drag forces, the $y$-direction far-field Gaussian wake shape 
in Eq. \ref{eq:uprimewake}
 suggests that in ALM, 
the drag force can be 
implemented using a kernel width $\epsilon_{d*}$ 
that can differ from that used for the lift force. 
Specifically, it can be chosen so as to 
mimic the initial width of the wake. 
At $x_* >>1$, the velocity defect 
integrates to $c ~U_{\infty} c_d/2$ as 
required by deficit momentum flux conservation. 
Thus  choosing $\epsilon_{d} = c~ c_d/2$, 
i.e.~the momentum thickness, should give 
realistic initial wake distributions. We remark that this choice implies that the
nonlinearity parameter is $u^\prime/U_\infty = 1/2\sqrt{\pi} \approx 0.28$, and so it is possible that  nonlinear effects 
begin to distort the velocity profile downstream when using such a kernel, but anyhow
downstream mixing and growth of the wake will 
naturally be accounted for by turbulence 
resolving or modeling parts of the LES. 

\begin{figure}[htb!]
	\centering
	\includegraphics[width=0.5\linewidth]
	{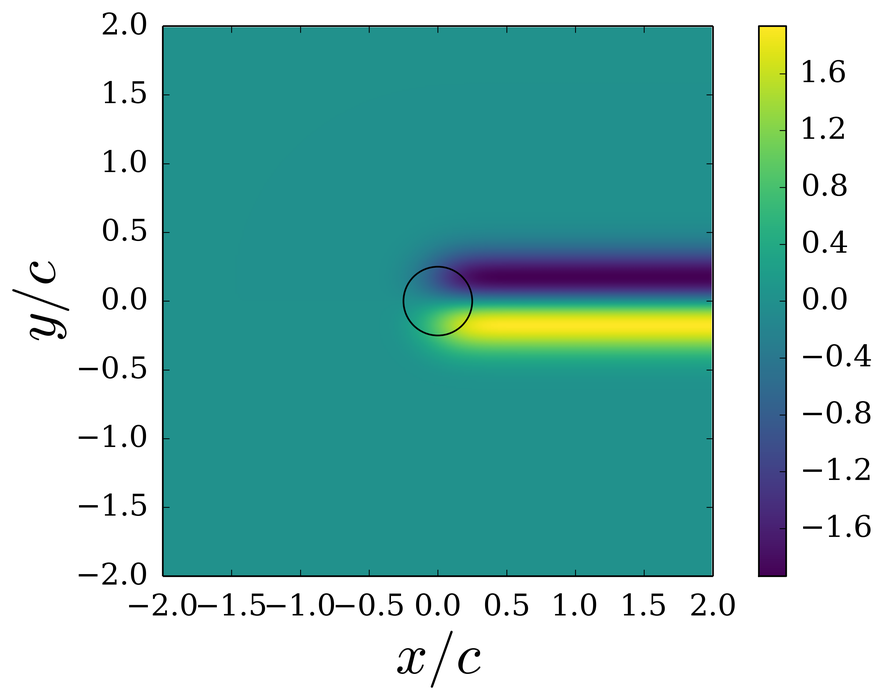}
	\caption{
	Vorticity perturbation distribution $\omega'_*$
		for a case with $\epsilon_{d*} = c_d/2$.
		The circle represents the 
		value of $\epsilon_{d*}$.
	}
	\label{fig:dragOmega}
\end{figure}

We remark that for instance highly stalled blades could 
be represented with larger $\epsilon_{d*}$  
than cases with initially very thin wakes. In time-dependent stall situations (e.g. in ALM simulations of vertical axis wind turbines), 
one may allow the kernel width to change in time. 
Figure \ref{fig:dragOmega} shows the vorticity distribution
for a drag force with $\epsilon_d$ chosen to be equal
to the momentum thickness.

\section{Generalization: 2D Gaussian Kernel}
\label{sec-2d}
The geometry of a typical airfoil is elongated in the $x$-axis and
much thinner in the $y$-axis.
For this reason
we now generalize the formulation presented for a 
circular Gaussian kernel
to an elliptical Gaussian kernel 
with different widths in the $x$- and $y$-directions. 
For simplicity we first consider the semi-axes 
aligned with x-y coordinate system:
\begin{equation}
\eta_\epsilon = \frac{1}{\epsilon_x \epsilon_y \pi^{2}} 
e^{-\left( x^2/\epsilon_x^2 + y^2/\epsilon_y^2 \right)}.
\end{equation}
The idea is to optimize the values of both 
$\epsilon_x$ and $\epsilon_y$ by, again, minimizing the error defined 
in equation \ref{eq:err}.
We seek a solution to the problem in a similar way to 
Section \ref{sec-solutionepsilon}.
After the linearization of the equations we obtain an expression
for the vorticity perturbation
\begin{equation}
	\omega_{\epsilon *}^\prime(x_*,y_*) = - 
	\frac{2 K_*}{\epsilon_{x*}\epsilon_{y*}} 
	e^{-(x_*^2/\epsilon_{x*}^2+y_*^2/\epsilon_{y*}^2)}.
    \label{eq:vorticity2D}
\end{equation}
The equation can be written in terms of the streamfunction
as the Poisson equation
\begin{equation}
	\nabla^2 \psi' = - \omega_{\epsilon *}^\prime(x_*,y_*) =
	\frac{2 K_*}{\epsilon_{x*}\epsilon_{y*}} 
	e^{-(x_*^2/\epsilon_{x*}^2+y_*^2/\epsilon_{y*}^2)}.
    \label{eq:streamF2D}
\end{equation}
The equation can be written in Fourier space
\begin{equation}
\widetilde{\psi'}(k_x, k_y)
= - \frac{
\widetilde{
\omega_{\epsilon *}^\prime}
(k_x, k_y)}
{k_x^2 + k_y^2}
= -
\frac{K_*
e^{-(k_x^2\epsilon_{x*}^2+k_y^2\epsilon_{y*}^2)/4}}
{ k_x^{2} + k_y^{2} }
.
\label{eq:PoiFS}
\end{equation}
The kernel can be rotated in space according to the angle of
attack by a simple transformation with
$x' = x \cos \alpha - y \cos \alpha$ and
$y' = x \sin \alpha + y \sin \alpha$.
The same transformation is done to the 
wave numbers $k_x$ and $k_y$ in Fourier Space
in order to rotate the solution by the given
angle of attack $\alpha$.
Eq. \ref{eq:PoiFS} is solved by using the 
Fast Fourier Transform 
algorithm \cite{cooley1965algorithm}
as implemented in the numpy library 
\cite{oliphant_python_2007}.
The solution is periodic, 
so artificial counter rotating vortices
are created near the edges
to artificially recover periodicity in the solution.
In order to overcome this, 
a solution for the  problem in Fourier Space
(Eqs. \ref{eq:PoiFS} with $\epsilon_{*y}=\epsilon_{*x}$)
is subtracted from the solution 
(Eq. \ref{eq:PoiFS}).
The solution
to the problem with $\epsilon_{*y}=\epsilon_{*x}$ 
(Eqs. \ref{eq:solutionueps}\textemdash\ref{eq:solutionveps})
is then added back in real space
to obtain the final form of the solution.
This method eliminates the 
opposing artificial circulation from the edges,
where the true solution must behave 
as an ideal vortex far away from the center
for both cases ($\epsilon_x = \epsilon_y$ and 
$\epsilon_x \ne \epsilon_y$).

\begin{figure}
\centering
\includegraphics[width=0.49\linewidth]
{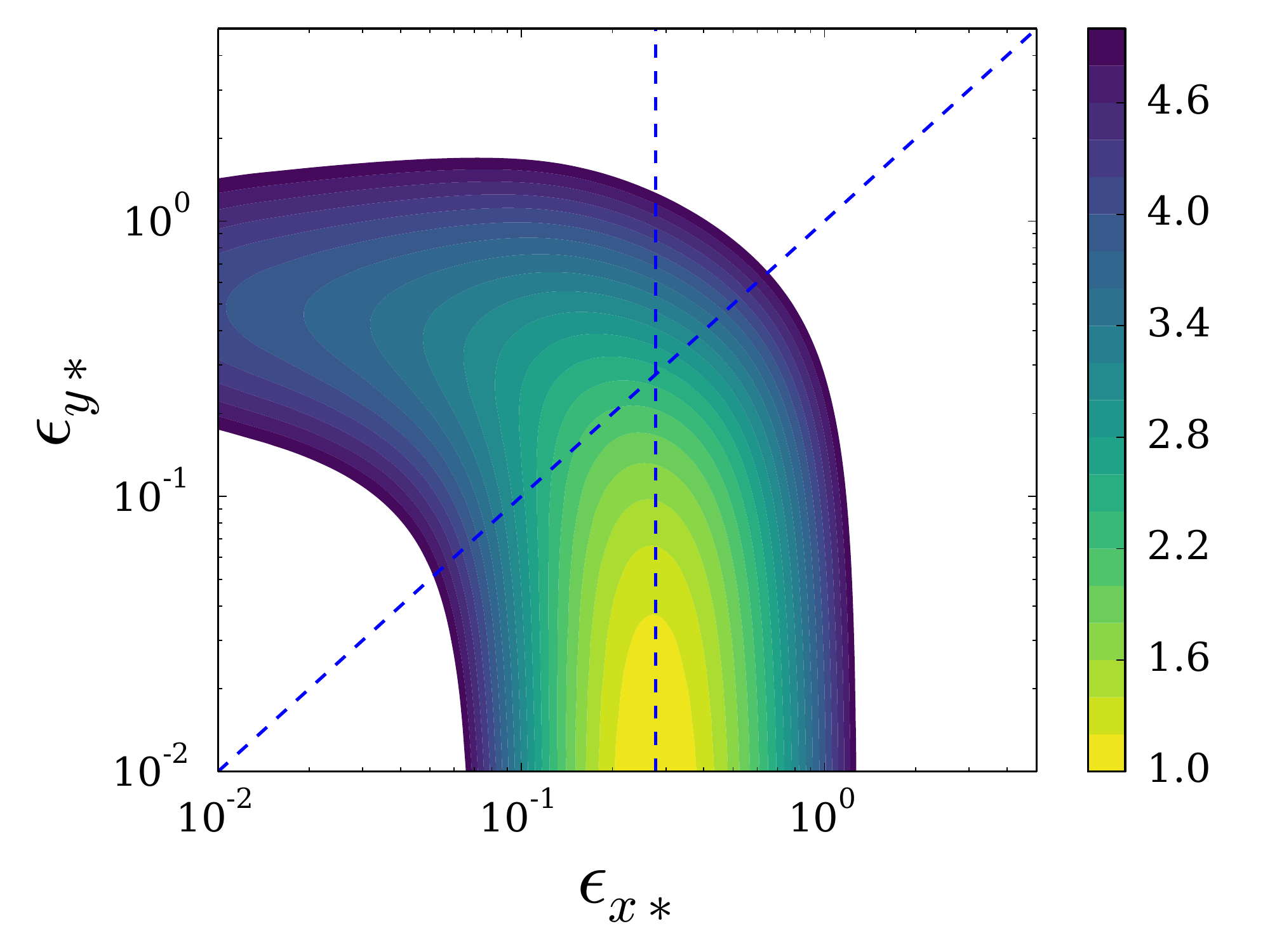}
\includegraphics[width=0.49\linewidth]
{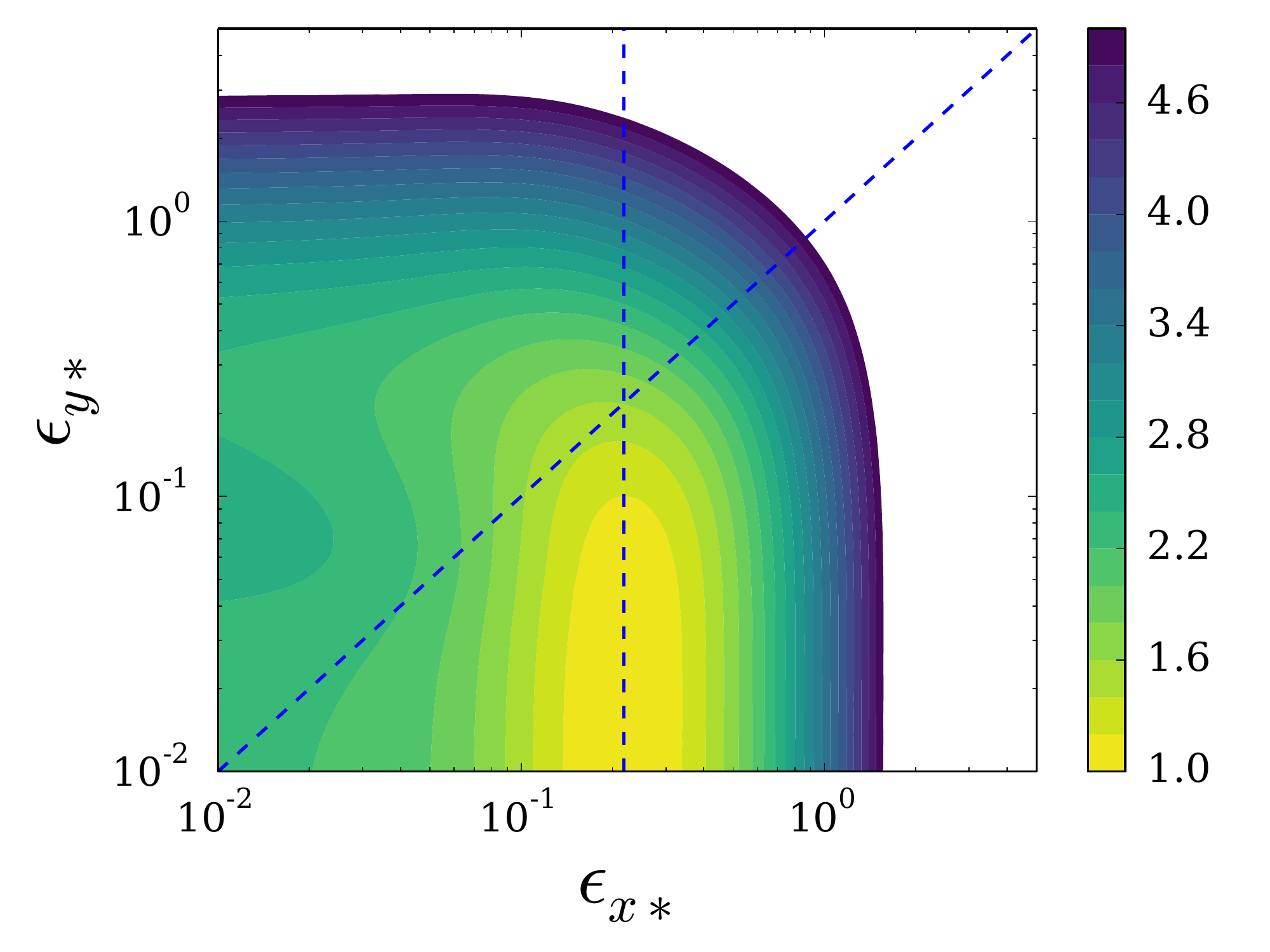}
\caption{ Normalized square error ($E_u^2/E_{u-min}^2$) 
contours as function of
Gaussian width $\epsilon_{x*}$ and $\epsilon_{y*}$
for the case with angle of attack $\alpha=12^o$. 
Results are shown for a flat plate (left) 
and for a symmetric Joukowski airfoil (right).
Vertical lines mark the optimal value of
$\epsilon_{x*}$, while the optimal value for
$\epsilon_{y*}$ tends to zero.
The diagonal line marks the case with
$\epsilon_{x*}=\epsilon_{y*}$.}
\label{fig:contourError2D}
\end{figure}

\begin{figure}
\centering
\includegraphics[width=0.49\linewidth]{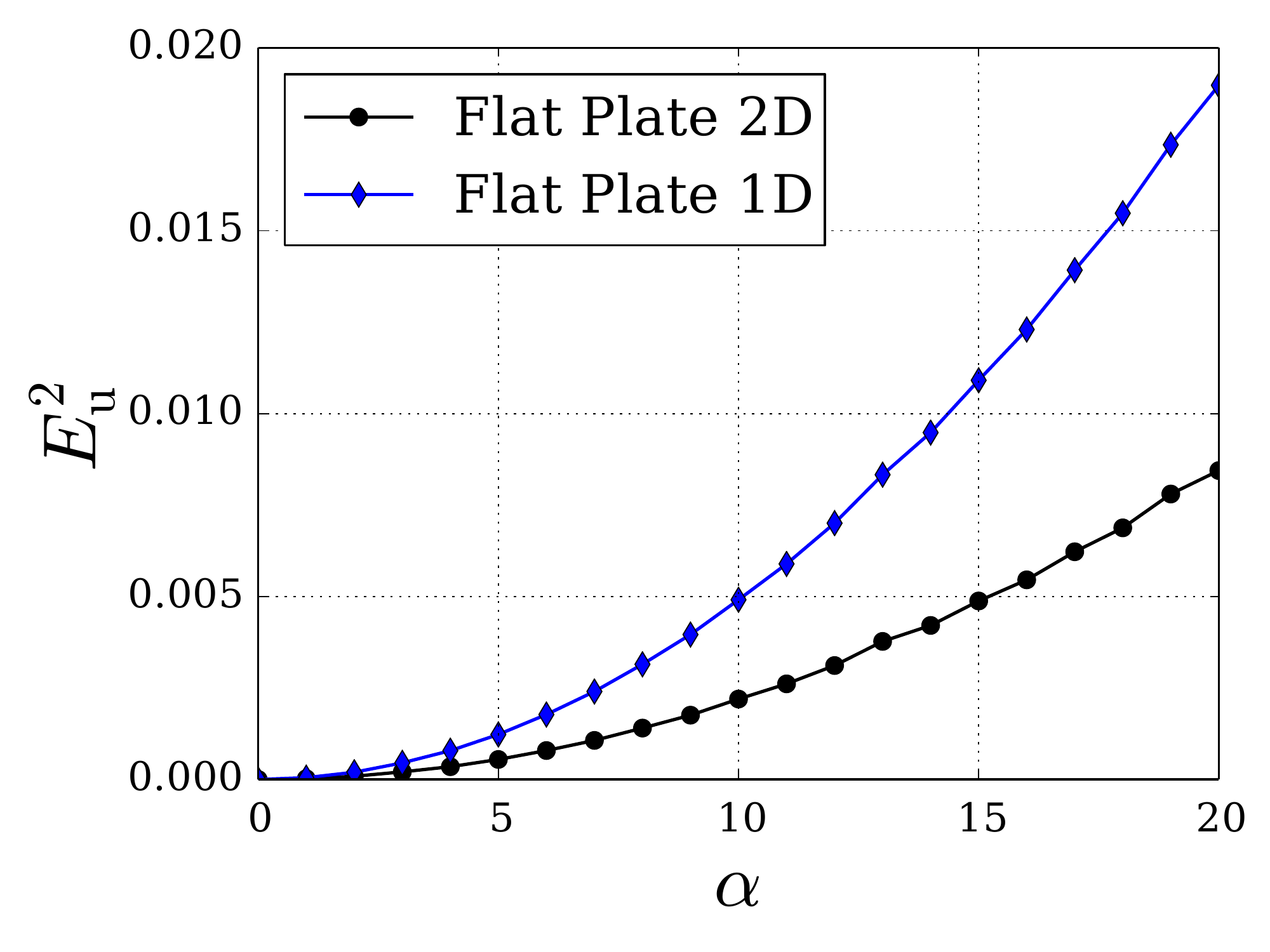}
\includegraphics[width=0.49\linewidth]{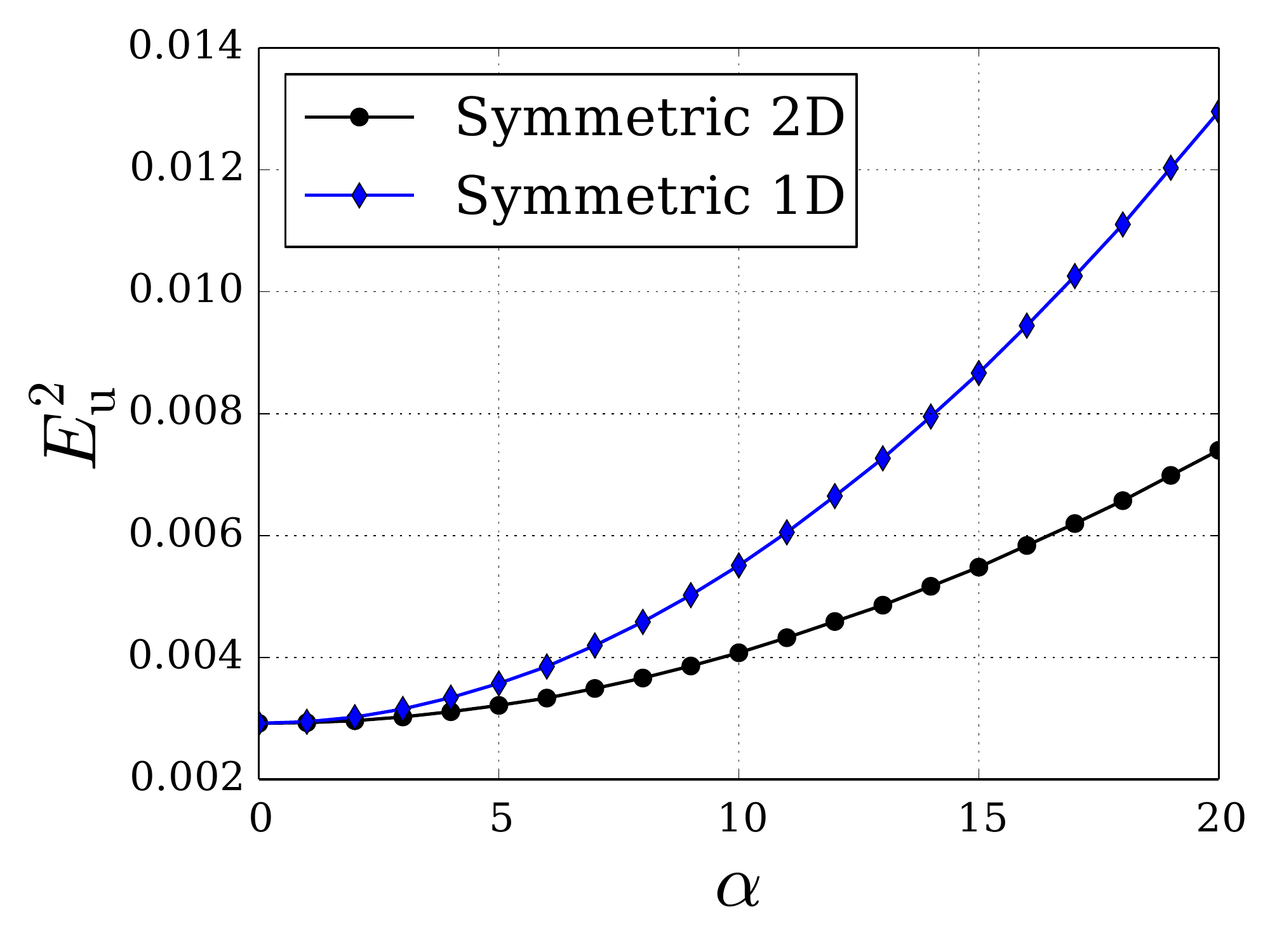}
\caption{ Squared error at the optimum values $\epsilon^{\rm opt}_{x*}$,
$\epsilon^{\rm opt}_{y*}$ 
and chord position $s_{*0}^{\rm opt}$ as a function of angle of attack $\alpha$ for a flat plate (left) and Symmetric airfoil (right).}
\label{fig:2DaoaError1D}
\end{figure}

The same optimization algorithm as in Section \ref{sec-optimization}
is used to find the optimum values for the kernel widths 
$\epsilon_{x*}$, $\epsilon_{y*}$ and the position $s_{0*}$.
Figure \ref{fig:contourError2D} shows contours 
of the normalized square error
as a function of $\epsilon_{x*}$ and $\epsilon_{y*}$.
It is seen that the optimum value is 
much more sensitive to
the width in the direction of the 
chord $\epsilon_{x*}$
than in the direction of 
thickness $\epsilon_{y*}$.
This method provides solutions with smaller errors
than the cases with $\epsilon_{x*}=\epsilon_{y*}$
shown in Section \ref{sec-optimization}.
The dashed diagonal line in Figure \ref{fig:contourError2D}
shows the case of $\epsilon_{x*}=\epsilon_{y*}$,
which lies above the optimal values for the 
elliptical case.
This line shows that even though
the optimal values for the case with
$\epsilon_{x*}\ne\epsilon_{y*}$ improve the
error as compared to the circular Gaussian,
the error is still within the same 
order of magnitude.
The difference in error magnitude 
is shown in Figure \ref{fig:2DaoaError1D}, where the
error is improved by more than 50\% for cases of a flat plate
and symmetric airfoil
for higher angles of attack.

\begin{figure}
	\centering
	\includegraphics[width=0.49\linewidth]
	{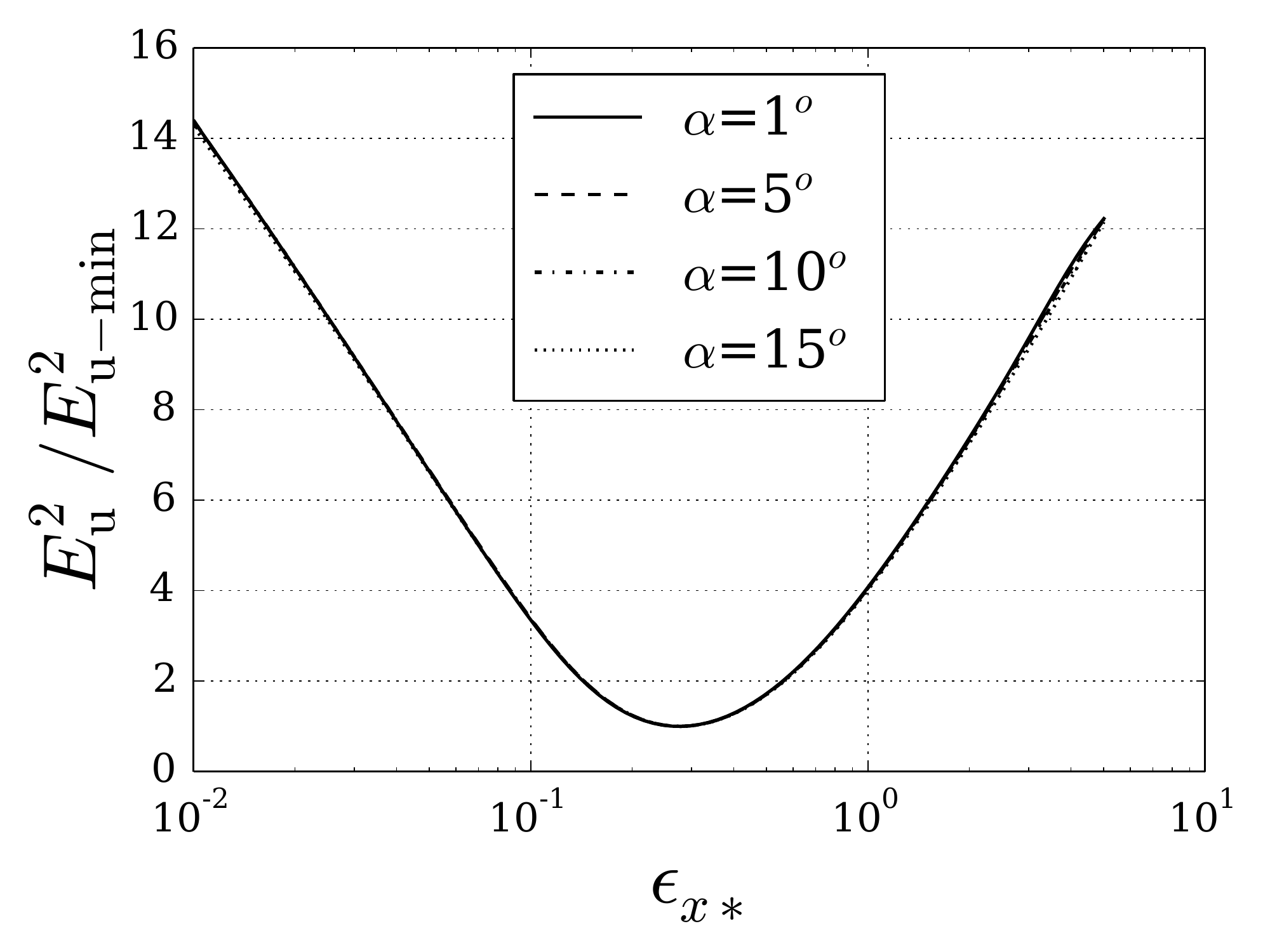}
	\includegraphics[width=0.49\linewidth]
	{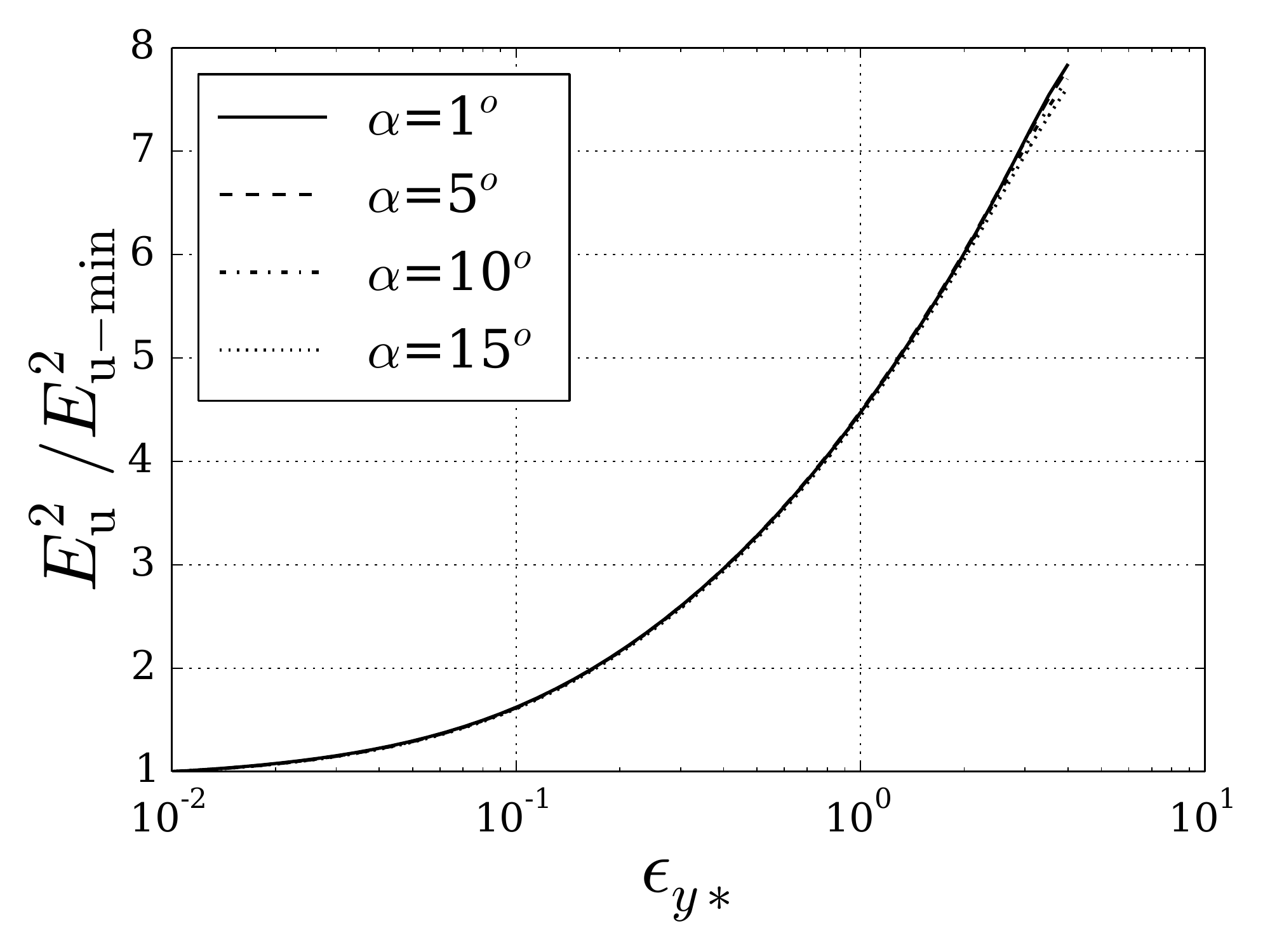}
	\caption{ Normalized squared error as a 
		function of $\epsilon_x$ (left) and
		$\epsilon_y$ (right) 
		for a flat plate at 
		different angles of attack.}
	\label{fig:1DError2D}
\end{figure}

\begin{figure}
 \centering
 \includegraphics[width=0.49\linewidth]
{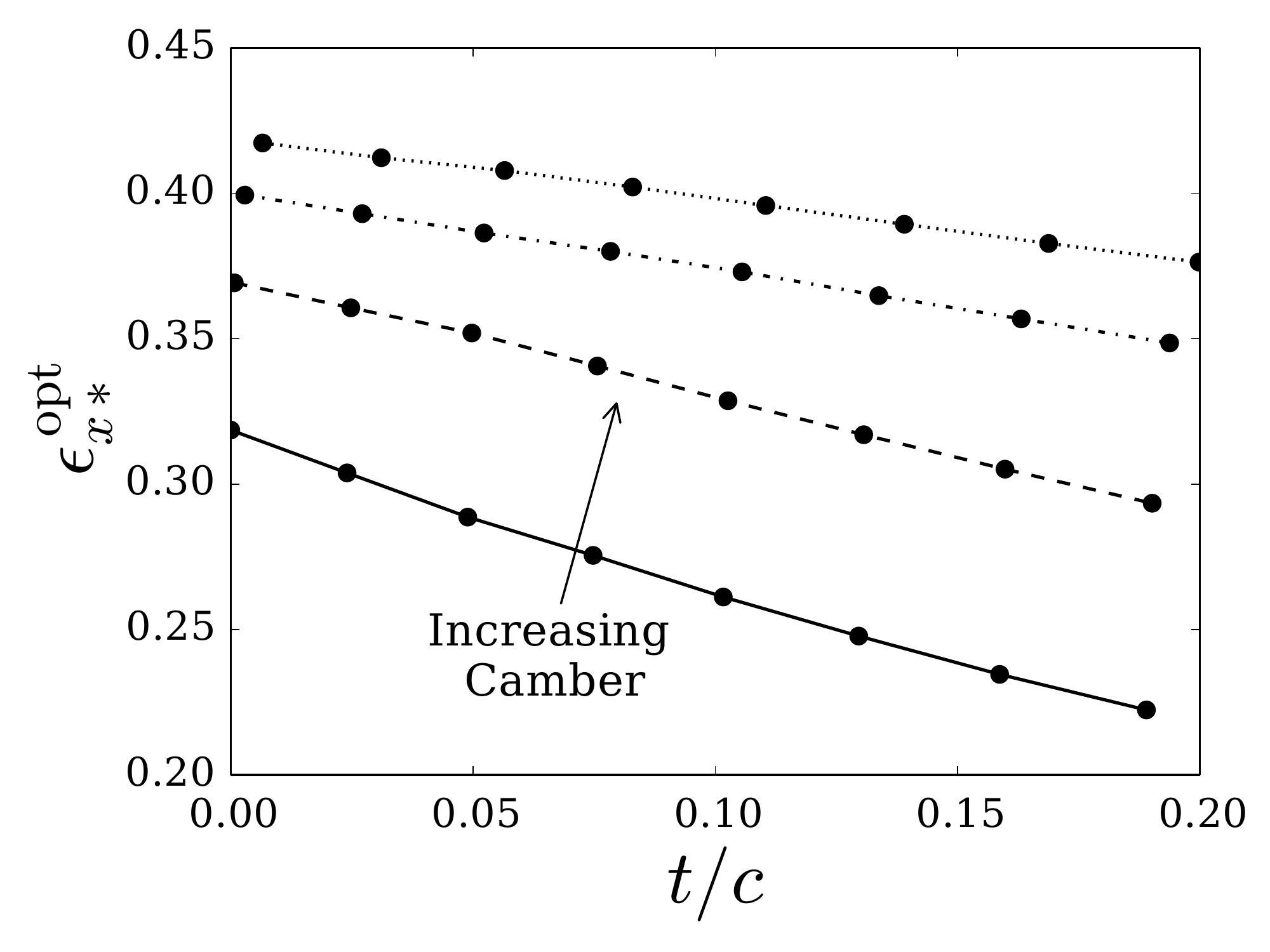}
 \includegraphics[width=0.49\linewidth]
{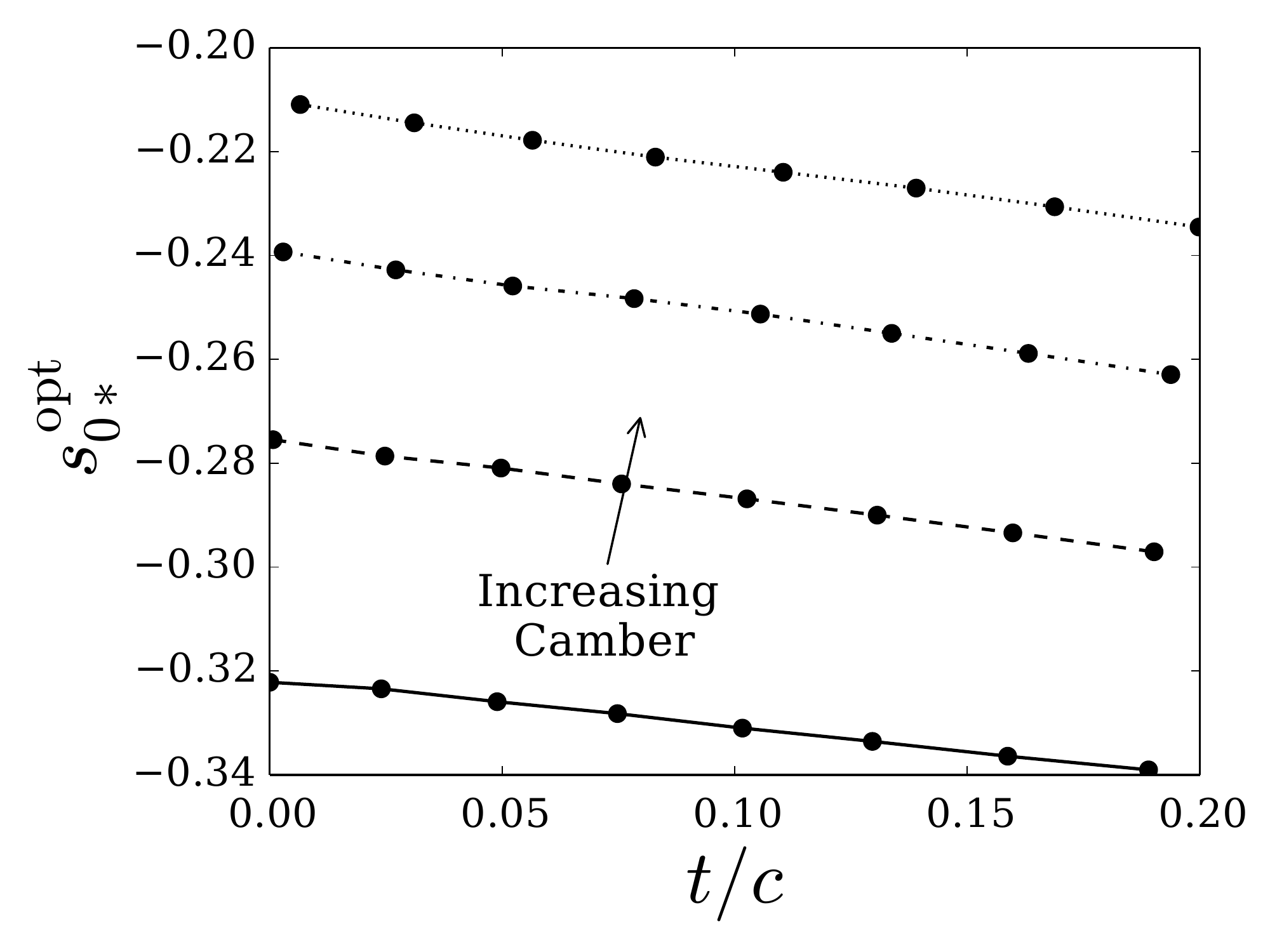}
 \caption{ 
 Optimum values $\epsilon^{\rm opt}_{x*}$ 
 and chord position $s_{*0}^{\rm opt}$ as 
 function of thickness $t/c$
 for different camber with angle of attack $\alpha=12^o$.}
 \label{fig:cambered2Dlines}
 \end{figure}

The optimum values are independent of angle of 
attack as shown in
Figure \ref{fig:1DError2D}.
The error is more sensitive to variation in $\epsilon_{*x}$,
after $\epsilon_{y*}$ reaches a value on the order of 
$\epsilon_{y*}\sim 10^{-2}$, the error reaches a threshold
and smaller values do not provide improved results.
This solution is similar to an actuator surface method,
but with a vorticity distribution along the chord 
given by the Gaussian 
field with an optimal $\epsilon_{x*}$.
In this case, the center location of the surface
is near the quarter chord
as shown previously in Section
\ref{sec-optimization}.
The kernel widths $\epsilon_{x*}$ and
$\epsilon_{y*}$ are always smaller
than the chord.
Figure \ref{fig:cambered2Dlines}
shows the distribution of $\epsilon^{\rm opt}_{x*}$ and
chord position $s_0^{\rm opt}$ for
different thicknesses and camber.
The values for $\epsilon^{\rm opt}_{x*}$ are 
larger than for the 1D case.
The optimal location $s_0^{\rm opt}$ is very similar
to the 1D case.

\begin{figure}
\centering
\includegraphics[width=0.32\linewidth]{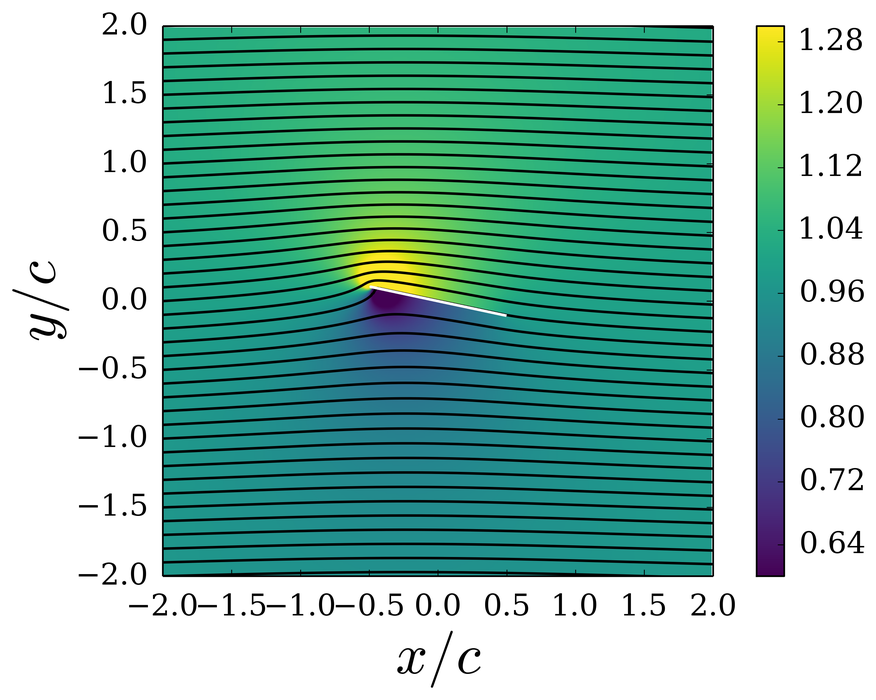}
\includegraphics[width=0.32\linewidth]{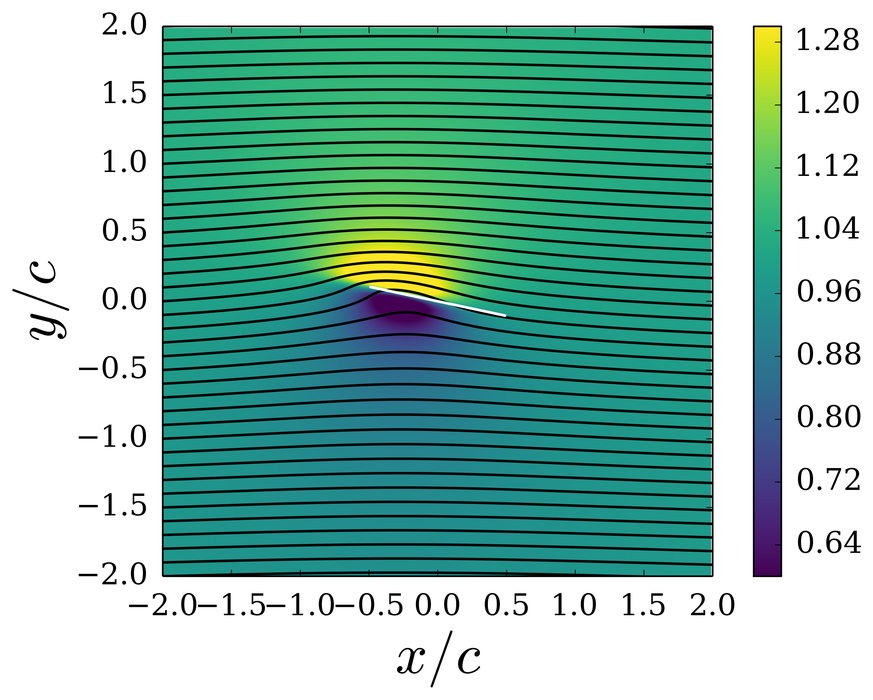}
\includegraphics[width=0.32\linewidth]{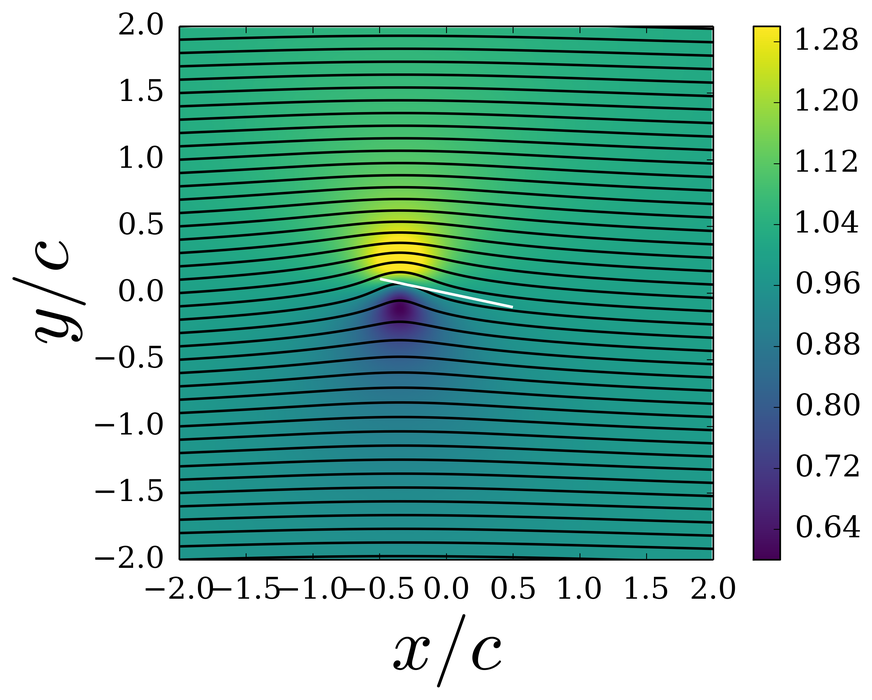}
\caption{ Visualization of potential flow solution over a Joukowski airfoil (left), body force solution with 
$\epsilon_{x*}$ and $\epsilon_{y*}$ (middle)
and body force solution with 1D Gaussian (right)
for a flat plate. Colors  indicate the magnitude of the velocity field while solid lines indicate streamlines.}
\label{fig:flatPlate2D}
\end{figure}

\begin{figure}
\centering
\includegraphics[width=0.32\linewidth]{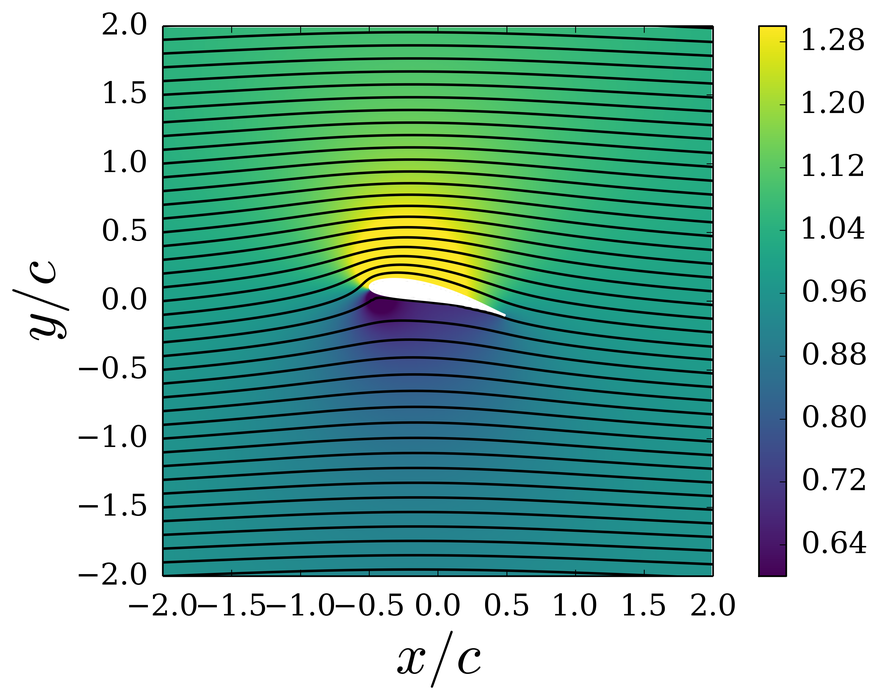}
\includegraphics[width=0.32\linewidth]{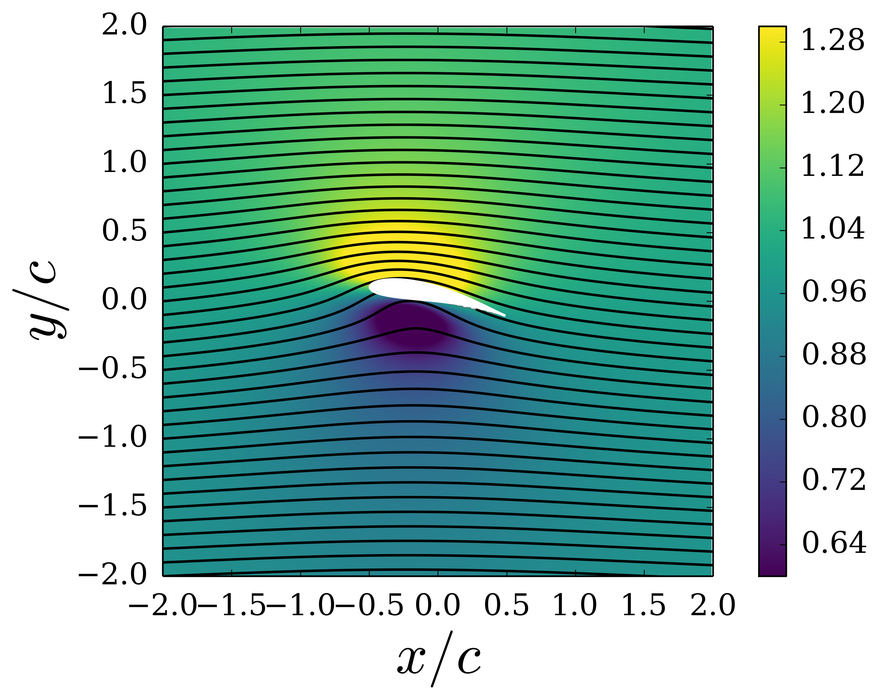}
\includegraphics[width=0.32\linewidth]{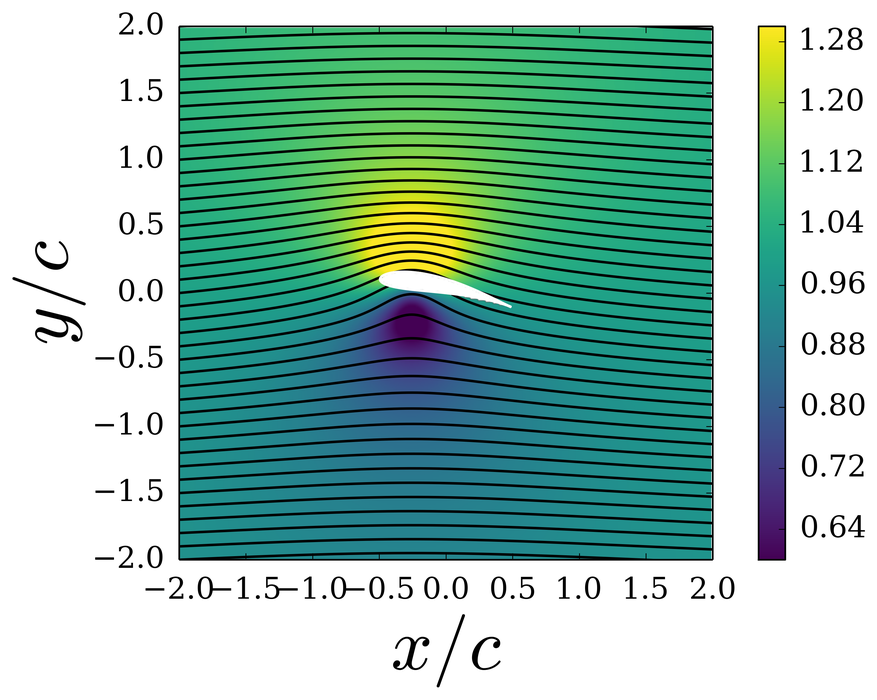}
\caption{Same as Figure \ref{fig:flatPlate2D} but for cambered airfoil case.}
\label{fig:Cambered2D}
\end{figure}

Figures \ref{fig:flatPlate2D} and \ref{fig:Cambered2D} show
velocity contours for the cases with optimal
$\epsilon_{x*}$ and $\epsilon_{y*}$.
The most noticeable feature of the
velocity field induced by the elliptical 2D
Gaussian lift force kernel
is that the deformation of the 
streamlines agrees better with
the potential flow solution
than the circular Gaussian kernel 
case considered in Section \ref{sec-solutionepsilon}.

\section{Conclusions}
By examining the velocity field induced by a 
circular Gaussian body force and 
comparing it with the velocity 
field one wishes to approximate 
(e.g. flow over an airfoil with uniform inflow), 
we can provide a criterion to 
select an optimum value of the 
force width $\epsilon$ and its 
position along the chord $s_{0}$. 
The analytical solution  
for the velocity field induced 
by a Gaussian body force is 
obtained using a perturbation 
analysis, i.e.~for a linearized 
advection velocity. 
Thus the analytical solution obtained 
is expected to become less accurate for large applied forces that cause significant velocity perturbations compared to $U_\infty$. 

We applied the method for both lift and drag forces.
First, the solutions were used to show that the common method of sampling the velocity at the center of 
the Gaussian provides the correct reference velocity due to the symmetric vorticity distribution that 
results from applying a Gausian lift force.  For the case of drag, 
a reference velocity correction factor was developed which depends on the ratio of the momentum 
thickness and kernel width used to specify the drag force.
 
Then solutions were used to determine the optimal kernel width $\epsilon$. 
In the case of a flat plate, to represent lift the values $\epsilon/c=0.17$ and $s_{0}/c = -0.36$ provide the optimum.
For the case of a symmetric Joukowski airfoil the values are within a range of $\epsilon/c=0.14$ to $0.17$
and $s_{0}/c = -0.37$ to $ -0.35$, essentially the same as those for the flat plate. We also found that these values do not depend strongly on angle of attack.
Further similar calculations for cambered airfoil provided very similar results but with a wider range of values from 
$\epsilon/c=0.14$ to $0.24$ and $s_{0}/c = -0.37$ to $ -0.24$ depending on camber and thickness.

The results from the theoretical analysis 
have the following practical implications: 
When using ALM with LES grid resolutions 
at or larger than the chord-length, 
the choice of ALM smoothing kernel 
scale must be dictated by LES 
numerical considerations, as is 
often done in practice
\citep{sorensen_numerical_2002,
	martinez-tossas_large_2014,
	jha2014guidelines}. 
Conversely, if the LES grid 
resolution is sufficiently 
refined to place a number of 
grid points along the chord 
of the lifting surface, then 
best results should be obtained 
when using a ``lift force'' 
Gaussian kernel with the physically 
optimal width and location determined 
in the present calculations. 
The results presented are only for Joukowski airfoils,
nevertheless, given the relative insensitivity of
$\epsilon_{\rm opt} / c$, for example, the flat plate 
and cambered airfoil cases
we expect the obtained $\epsilon_{\rm opt} / c \sim 0.2$
value to be approximately valid for other types of airfoils.
Values close to the optimal should provide smaller
errors when trying to replicate the flow field of
a 2D airfoil.
A 2D elliptical Gaussian kernel 
with a width $\epsilon_{x*}$ 
in the direction of the chord and $\epsilon_{y*}$ 
in the thickness direction
can provide even more accurate
results than a circular Gaussian kernel.
The velocity error for this kernel is further reduced 
compared to the 1D kernel.
The position of this 2D kernel in the chord is similar
to the 1D kernel $s_{0}/c$.
For the drag force, a separate Gaussian 
force with kernel width that scales with the 
wake momentum thickness could be used to generate
a velocity wake profile with a
realistic initial thickness.
The results shown are for 2D lift
and drag forces.
Future work includes testing these optimal values
in the case of a 3D rotor, with special interest in
the tips of a rotating blade.

\acks
The authors thank Xiang I.A. Yang, Perry Johnson and
Philippe Spalart for a number of important ideas and suggestions at various stages of this work.
The research was supported by the National Science Foundation grants no. IGERT 0801471,
1243482 (the WINDINSPIRE project) and 1230788.
Numerical implementation and plots were generated using python with scipy, numpy and matplotlib libraries. 

\bibliographystyle{unsrt}
\bibliography{bibliography}

\end{document}